\documentclass[a4paper,noarxiv]{quantumarticle}
\usepackage{amsmath,amsthm,amssymb,amsfonts,graphicx}
\usepackage{braket}
\usepackage{calc}
\usepackage{bbold}
\usepackage[numbers,sort&compress]{natbib}

\begin{document}

\title{Quantum Consensus Dynamics by Entangling Maxwell Demon}

\author{Sungguen Ryu}
\email{sungguen@ifisc.uib-csic.es}
\affiliation{Instituto de F\'{i}sica Interdisciplinar y Sistemas Complejos IFISC (CSIC-UIB), E-07122 Palma de Mallorca, Spain}

\author{Rosa L\'{o}pez}
\email{rosa.lopez-gonzalo@uib.es}
\affiliation{Instituto de F\'{i}sica Interdisciplinar y Sistemas Complejos IFISC (CSIC-UIB), E-07122 Palma de Mallorca, Spain}

\author{Ra\'{u}l Toral}
\email{raul@ifisc.uib-csic.es}
\affiliation{Instituto de F\'{i}sica Interdisciplinar y Sistemas Complejos IFISC (CSIC-UIB), E-07122 Palma de Mallorca, Spain}

\date{\today}

\begin{abstract}
 We introduce a Maxwell demon which generates many-body entanglement robustly against bit-flip noises, which allows us to obtain quantum advantage.
 Adopting the protocol of the voter model used for opinion dynamics approaching consensus, the demon randomly selects a qubit pair and performs a quantum feedback control, in continuous repetitions.
 We derive upper bounds of the entropy reduction and the work extraction rates by demon's operation, which are determined by a competition between the quantum-classical mutual information acquired by the demon and the absolute irreversibility of the feedback control.
 Our finding of the upper bounds corresponds to a reformulation of the second law of thermodynamics under a class of Maxwell demon which generates many-body entanglement in a working substance.
\end{abstract}

\maketitle

\section{Introduction} 

A modern view of a Maxwell demon is based on a closed-loop feedback control in which the dynamics of a system uses outputs of a measurement as inputs in a smart manner. Within the framework of nanoscale machines, quantum feedback controls involve systematic measurements and manipulation of quantum systems with the aim of extracting useful work or cooling a system~\cite{kim2011quantum,koski2015chip,pekola2015towards,funo2015quantum,koski2014experimental,liuzzo2016thermodynamics}.
The field of thermodynamics of information~\cite{sagawa2008second,parrondo2015thermodynamics,funo2015quantum, ptaszynski2019thermodynamics} has clarified the fundamental bounds on entropy reduction and work extraction in terms of quantities such as the quantum-classical mutual information and absolute irreversibility~\cite{funo2015quantum}. However, the thermodynamics of continuous quantum feedback~\cite{ptaszynski2019thermodynamics} remains a largely unexplored issue. Recently, variants of the original Maxwell demon which operate continuously in time have been demonstrated~\cite{manzano2021thermodynamics,ribezzi2019large}, showing an enhancement of the work extraction beyond the conventional feedback control with a limitation given by modified second-law-like inequalities.

\begin{figure}[t]
 \centering
 \includegraphics[width=\columnwidth]{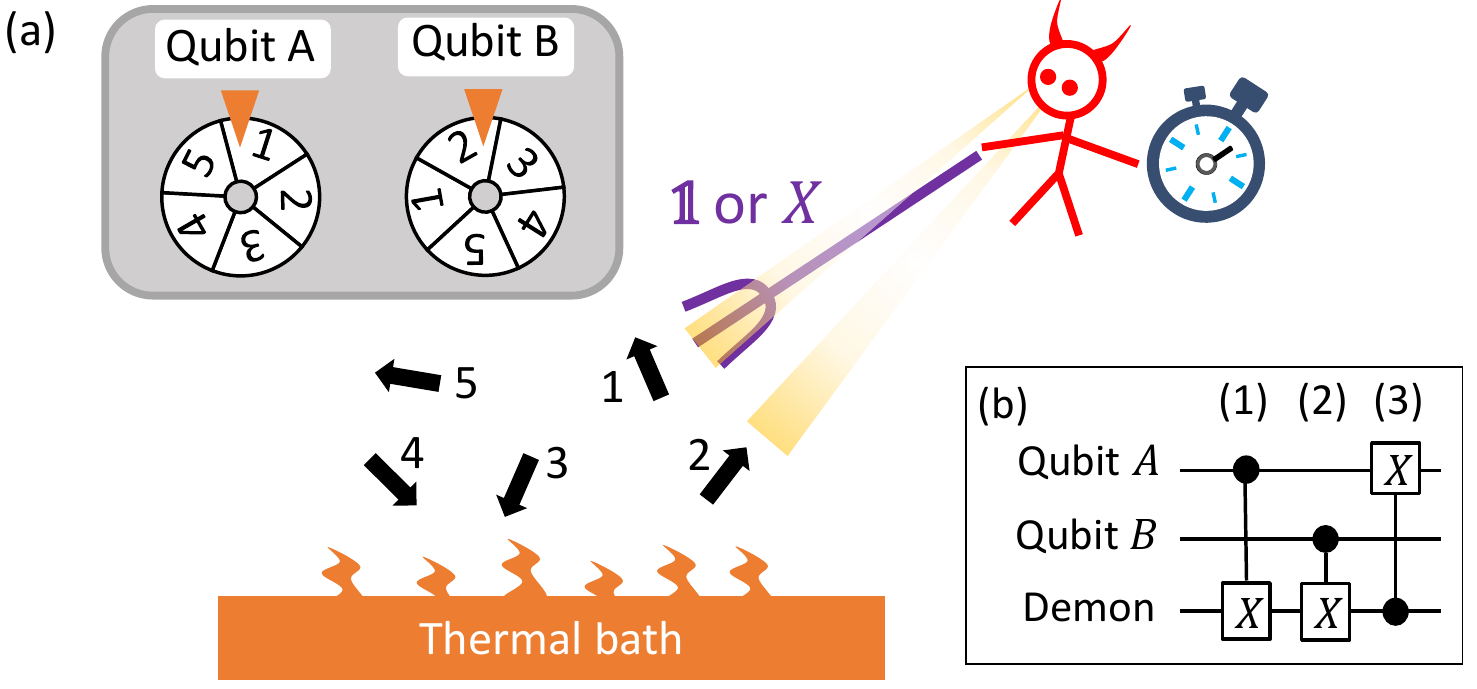}
 \caption{Entangling Maxwell demon adopting a protocol of the voter model.
(a) The demon selects a random pair of qubit A and B among many, e.g. using a roulette. Then, it induces a quantum feedback control to the selected qubits, a copy process in this case.
In the copy process, the demon first measures whether the two qubits are in the same basis state or not, then flips the qubit A only for the latter case. The whole process from the selection to the copy is repeated with a rate $\Gamma_{\text{copy}}$. A thermal bath induces a bit-flip with a rate $\Gamma_{\text{flip}}$.
 (b) The quantum feedback control realizing the copy process. It is series of controlled-NOT gates [(1)-(3)] where the black circles depict the control and rectangular boxes depict the target.
}
 \label{fig:setup}
 \end{figure}

Stepping further along this direction, we propose in this paper a new type of Maxwell demon, namely a continuous quantum feedback control, that is capable of generating many-body entanglement in the working substance. Our demon (see Fig.~\ref{fig:setup}) acts by randomly selecting two qubits A and B among many, and inducing a quantum feedback control on them which reduces entropy and enhances correlation simultaneously. 
The demon continuously repeats the selection and the feedback control.

In fact, such protocol realizes the quantum steady-state engineering~\cite{verstraete2009quantum,ticozzi2012stabilizing,ticozzi2014steady,wiseman2009quantum,zhang2017quantum,cho2011optical,lin2013dissipative,reiter2016scalable,stevenson2011engineering,feng2011generating,morigi2015dissipative} as studied in quantum information and optics.
The two-qubit quantum feedback control realizes the two-particle dissipations~\cite{verstraete2009quantum, ticozzi2012stabilizing,ticozzi2014steady,diehl2008quantum,wintermantel2020unitary} whereby a dissipation on a particle depends on the state of the other. 
Previous studies have identified possible types of entangled states which are stabilizable.
However, the quantum dynamics, i.e. how entanglement, coherence, and von Neumann entropy evolve in time, still lacks understanding.
The mechanism behind the quantum dynamics is nontrivial due to two simultaneous tasks, the random selection and the continuous quantum measurement~\cite{Belenchia:2020,elouard2018efficient}.
Moreover, the fundamental bound of entropy reduction rate by the two-particle dissipations has not been studied before, and an identification of the bound requires a viewpoint from thermodynamics of information.

In this paper, we study the quantum dynamics and the second law of thermodynamics under the action of the entangling Maxwell demon.
We first propose a quantum version of the voter model, an entangling Maxwell demon adopting a protocol inspired by the noisy voter model~\cite{Clifford:1973,Holley:1975,Granovsky:1995,Castellano:2009,Peralta_pair:2018}, and motivated by the fact that the classical model generates classical correlation of human opinions among agents. 
Our main finding is that Greenberger-Horne-Zeilinger (GHZ) entanglement~\cite{Leibfried:2005,monz201114,giovannetti2006quantum} is generated among the working substance and stabilized against the bit-flip noises~\cite{dur2014improved,chaves2013noisy,wasilewski2010quantum}.
During the entanglement generation, the purity and the entropy of the working substance change non-monotonically in time, which turns out to be due the competition of the information gain and the absolute irreversibility of the feedback control.

Then, we reformulate the second law of thermodynamics under the action of a generic class of entangling Maxwell demons.
We derive an upper bound of the entropy reduction, or equivalently an upper bound of the work extraction. The bounds are determined by the competition of the quantum-classical mutual information acquired by the demon and the associated absolute irreversibility of the control. We recall that the absolute irreversibility is defined as the sum of probabilities of the backward trajectories which do not have a corresponding forward trajectory, being responsible, for instance, of the breakdown of the Jarzynski equality in the free gas expansion~\cite{murashita2014nonequilibrium, murashita2017gibbs}. In our case, the absolute irreversibility is generally non vanishing as the quantum measurement in the action of the entangling Maxwell demon projects the state of the working substance to a local subspace which depends on the selection, an inevitable factor hindering the entanglement generation or work extraction. Our findings provide a necessary condition, namely that the information gain should be larger than the absolute irreversibility, to determine when an entangling Maxwell demon can successfully stabilize the many-body entanglement or extract work through the entangled working substance.

This paper is structured as follows.
In Sec.~\ref{sec:model} we present the details of the quantum voter model and the master equation in the Lindblad form.
In Sec.~\ref{sec:ent} we present the results of the quantum dynamics and discuss a possible method for the experimental realization.
In Sec.~\ref{sec:ubs}, we derive the upper bounds of the entropy reduction and the work extraction rates by the entangling Maxwell demon.
In Sec.~\ref{sec:lambda} we give an interpretation of the absolute irreversibility which appears in the bounds.
In Sec.~\ref{sec:comp} we show that the non-monotonic behavior in the purity and the entropy of the quantum voter model is due to the competition of the absolute irreversibility and the information gain of the feedback control.
In Sec.~\ref{sec:egDemon}, an example of the work extraction by the entangling Maxwell demon is shown. Finally, in Sec.~\ref{sec:conclusion} we summarize our main results.

\section{Quantum Voter Model}
\label{sec:QV }
\subsection{Model}
\label{sec:model}
In the following, we explain the quantum version of the voter model. Let us consider a system of $N$ qubits ($N\ge 2$). In our analogy, the two level state $|s_i\rangle$ ($\ket{0}$ or $\ket{1}$) of qubit $i=1,\dots,N$, plays the role of the opinion variable $s_i=0,1$ hold by an agent of the classical voter model, see Fig.~\ref{fig:setup}(a). A quantum feedback control (see below) realizes the copy process $C_{i \leftarrow j}$ whereby qubit $i$ copies the state of qubit $j$. We focus on the case of all-to-all connectivity, so that a copy process $C_{i \leftarrow j}$ of randomly chosen $i$ and $j \neq i$ is induced with rate $\Gamma_{\text{copy}}$.
(See Appendix~\ref{apx:1Dnn} for comparison with one-dimensional nearest-neighbor connections.)
A thermal bath around the system induces bit-flip noise, and the state $|s_i\rangle$ of the qubit $i$ flips ($|0\rangle \rightarrow |1\rangle$ or $|1\rangle \rightarrow |0\rangle$) with a rate $\Gamma_{\text{flip}}$ for random $i$.
Such flip process corresponds to the transversal noise~\cite{dur2014improved,chaves2013noisy,wasilewski2010quantum} induced by the dephasing in the basis of $\ket{0}\pm \ket{1}$.
Note that as discussed in Sec.~\ref{sec:egDemon}, $\Gamma_{\text{flip}}$ is determined by the temperature of the bath~\cite{schaller2014open}.
When there is no coherence, our system corresponds to the voter model when $\Gamma_{\text{flip}}=0$ and to the noisy voter model when $\Gamma_{\text{flip}}>0$~\cite{Peralta_pair:2018}.

The copy process $\mathcal{C}_{i \leftarrow j}$ cannot be realized only by unitary dynamics among the $N$ qubits because of the redundancy of the outcomes, e.g. $(s_i,s_j)=(0,0)$ and $(1,0)$ both become $(0,0)$ after the copy process. This process $\mathcal{C}_{i \leftarrow j}$ can be realized instead using a quantum feedback control with an ancilla qubit (which plays the role of a Maxwell demon) effectively measuring whether $s_i = s_j$ or not, see Fig.~\ref{fig:setup} (b). 
The joint state of qubits $i$, $j$, and ancilla, $\ket{s_i s_j s_a}$, changes by the three controlled-NOT (C-NOT) gates as
 \begin{equation}
 \label{eq:U12a}
 \begin{aligned}
 \ket{000} &\stackrel{(1)}{\rightarrow} \ket{000}
 \stackrel{(2)}{\rightarrow} \ket{000}
 \stackrel{(3)}{\rightarrow} \ket{000} \\
 \ket{010} &\rightarrow \ket{010} \rightarrow \ket{011} \rightarrow \ket{111} \\
 \ket{100} &\rightarrow \ket{101} \rightarrow \ket{101} \rightarrow \ket{001} \\
 \ket{110} &\rightarrow \ket{111} \rightarrow \ket{110} \rightarrow \ket{110}
\end{aligned}
\end{equation}
The first two gates, steps (1) and (2), effectively measure whether $s_i = s_j$ by flipping the ancilla qubit $s_a=0$ to $s_a=1$ only when $s_i \neq s_j$, while the third gate, step (3), realizes the copy process.
Note that this copy process is not prohibited by the no-cloning theorem~\cite{nielson2006quantum} because it does not copy an arbitrary state. The ancilla state should be initialized to $s_a = 0$ before starting the copy process, otherwise we obtain the opposite result, i.e. $s_i\neq s_j$ after the feedback. We assume an ideal ancilla in this study, which can be realized by attaching a separate bath~\cite{koski2015chip} that relaxes the ancilla into the state $\ket{0}$ in a time scale much faster than $\Gamma_{\text{copy}}^{-1}$. Below, we focus on the dynamics of the $N$ qubits, tracing out the ancilla.

The copy process $\mathcal{C}_{i\leftarrow j}$ changes the density matrix $\rho$ of the $N$-qubit state to $\mathcal{C}_{i \leftarrow j}(\rho)$,
\begin{equation}
 \label{eq:Cij}
 \mathcal{C}_{i \leftarrow j}(\rho)
 = \sum_{k=0,1} U_k^{(i)} M_k^{(ij)} \rho M_k^{(ij)}{U_k^{(i)\dagger}}.
\end{equation}
Here, $M_{0}^{(ij)}$ (resp. $M_{1}^{(ij)}$) is the measurement operator for the outcome $s_i=s_j$ (resp. $s_i \neq s_j$),
\begin{align}
 M_{0}^{(ij)} &\equiv \ket{0_i 0_j}\bra{0_i 0_j} + \ket{1_i 1_j} \bra{1_i 1_j} , \\
 M_{1}^{(ij)} &\equiv \ket{0_i 1_j}\bra{0_i 1_j} +\ket{1_i 0_j}\bra{1_i 0_j} .
\end{align}
The probability of each one of these outcomes is
\begin{equation}
 \label{eq:2}
 p_k^{(ij)} = \text{Tr} (M_k^{(ij)} \rho M_k^{(ij)}) , \text{ for } k=0,1 .
\end{equation}
The post-measurement state is
\begin{equation}
 \label{eq:rho-k}
 \rho_k^{(ij)} \equiv \frac{1}{p_k^{(ij)} } M_k^{(ij)} \rho M_k^{(ij)}.
\end{equation}
$U_k^{(i)}$ is the feedback operation for each measurement outcome,
\begin{equation}
 U_0^{(i)} =\mathbb{1} , \qquad
 U_1^{(i)}=X_{i} .
\end{equation}
where $X_i$ is the operator flipping the state of the qubit $i$: $X_i \ket{0_i}=\ket{1_i}$ and $X_i \ket{1_i}= \ket{0_i}$. Note that $U_k^{(i)}$ is a unitary and Hermitian operator for the copy process.

\begin{figure}[t]
 \centering \includegraphics[width=\columnwidth]{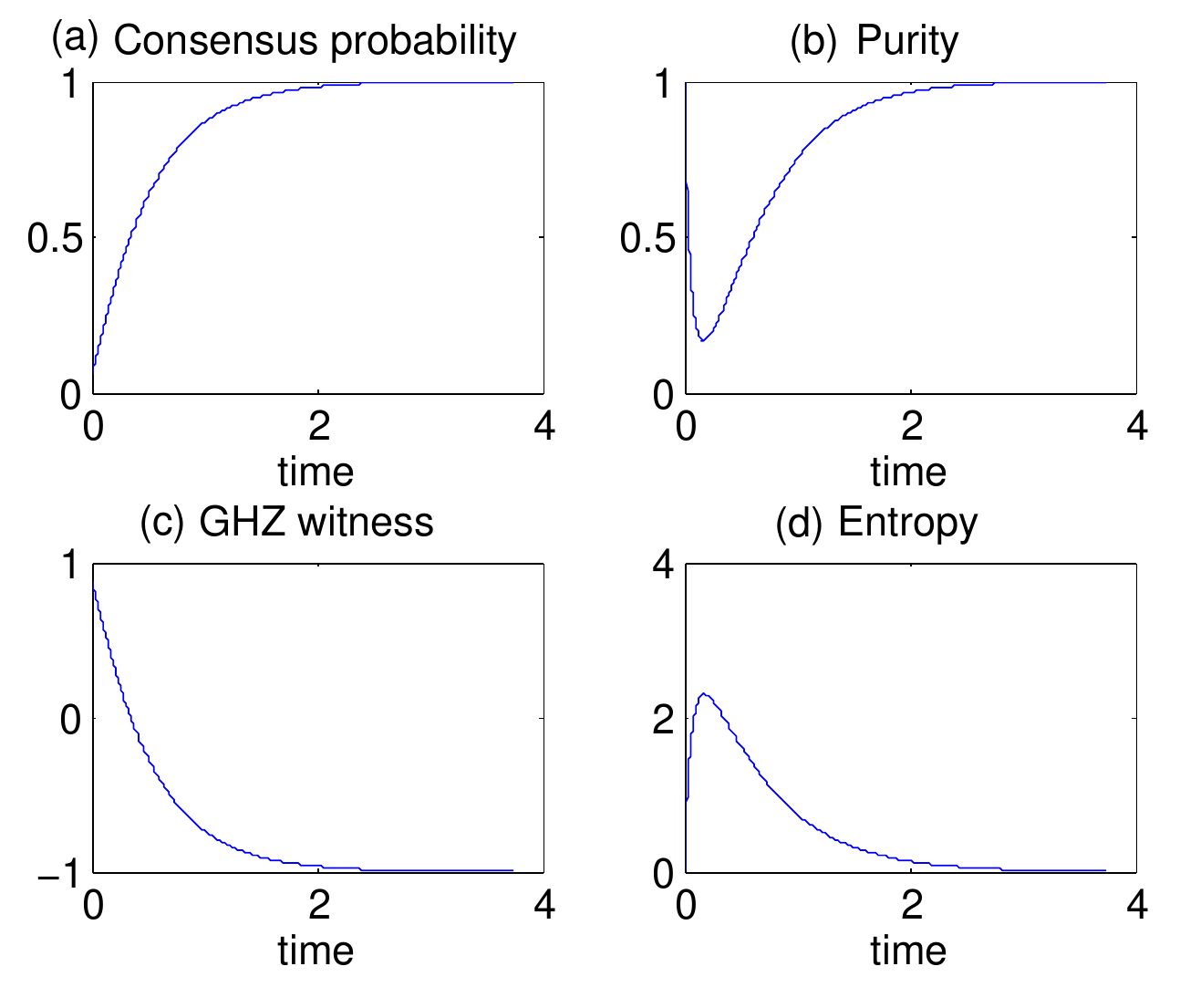}
 \caption{Entanglement generation by the consensus dynamics in the absence of bit-flip noise.
 The quantum feedback controls induce consensus among the qubits (a), while enhancing GHZ entanglement (c).
 They reduce [or increase] entropy of the qubits depending on time (d), accompanying the enhancement [or reduction] of coherence (b).
 Here $N=5$, $\Gamma_{\text{flip}}=0$, and the initial state is $\sum_{s_1,\cdots,s_N =0,1} \ket{s_1 \cdots s_N}$. The time is measured in $\Gamma_{\text{copy}}^{-1}$. }
 \label{fig:ghzN}
 \end{figure}

The time evolution of $\rho$ by the combined effect of the random copy processes and flips is given by 
\begin{equation}
 \label{eq:rho_C_N}
 \dot{\rho}= \Gamma_{\text{copy}} \sum_{\substack{i,j=1\\i\neq j}}^N \left( \mathcal{C}_{i \leftarrow j} (\rho) -\rho \right)
 + \Gamma_{\text{flip}} \sum_{i=1}^N \left( X_i \rho X_i^\dagger -\rho\right) .
\end{equation}
Equation (\ref{eq:rho_C_N}) is derived by considering a completely positive map of a small-time evolution, hence it is in the Lindblad form~\cite{wiseman2009quantum,schaller2014open}; see Appendix~\ref{apx:Lindblad-X} for details.
To obtain the results of Figs.~\ref{fig:ghzN}--\ref{fig:S-I} described below, the time evolution of $\rho$ has been calculated by a numerical integration of Eq.~(\ref{eq:rho_C_N}). For brevity in the notation, we define the total number of copy processes $C_{i \leftarrow j}$ for different choices of $i$ and $j$, as $N_{\text{copy}}\equiv N(N-1)$.

\subsection{Entanglement generation and stabilization}
\label{sec:ent}

To understand the effect generated by the copy process dynamics, we first focus on the case when there is no bit-flip noise, $\Gamma_{\text{flip}}=0$, a purely {\slshape consensus dynamics} in the language of the voter model~\cite{Castellano:2009}. 

Using as an initial state a symmetric superposition of all possible opinion configurations,
\begin{equation}
 \label{eq:4}
 \ket{\psi} = \sum_{s_1,\cdots,s_N =1,2} \ket{s_1 \cdots s_N}
\end{equation}
(here and henceforth we omit the normalization factor of all the pure states), 
the different panels of Fig.~\ref{fig:ghzN} provide evidence that this dynamics generates GHZ entanglement. Panel (a) shows that the consensus probability 
\begin{equation}
P_c(\rho) \equiv \bra{0,\dots,0}\rho\ket{0,\cdots 0}+\bra{1,\dots,1}\rho\ket{1,\cdots 1}
\end{equation}
increases in time and converges to $1$. When approaching the consensus, the state tends to the GHZ state, $\ket{0 \cdots 0} + \ket{1 \cdots 1}$, as evidentiated by the GHZ-witness value, defined as~\cite{monz201114},
\begin{equation}
 \label{eq:5}
 W_\text{GHZ} (\rho)= 1 - P_c(\rho)-2|\bra{0,\dots,0}\rho\ket{1,\dots,1}| ,
\end{equation}
displayed in panel (c).
The negative value of $W_\text{GHZ}$ means that the state displays a GHZ entanglement, where the lower bound $-1$ is achieved for the GHZ state~\cite{Leibfried:2005}. Interestingly, as displayed in panel (d), the von Neumann entropy,
\begin{equation}
 \label{eq:6}
 S(\rho)=-\text{Tr} (\rho\log\rho) ,
\end{equation}
initially increases and then decreases, while the purity, $\text{Tr} (\rho^2 )$, initially decreases and then increases, see panel (b). As discussed in Sec.~\ref{sec:comp}, this turns out to be the result of a competition between the quantum-classical mutual information and the absolute irreversibility.

 \begin{figure}[t]
 \centering \includegraphics[width=\columnwidth]{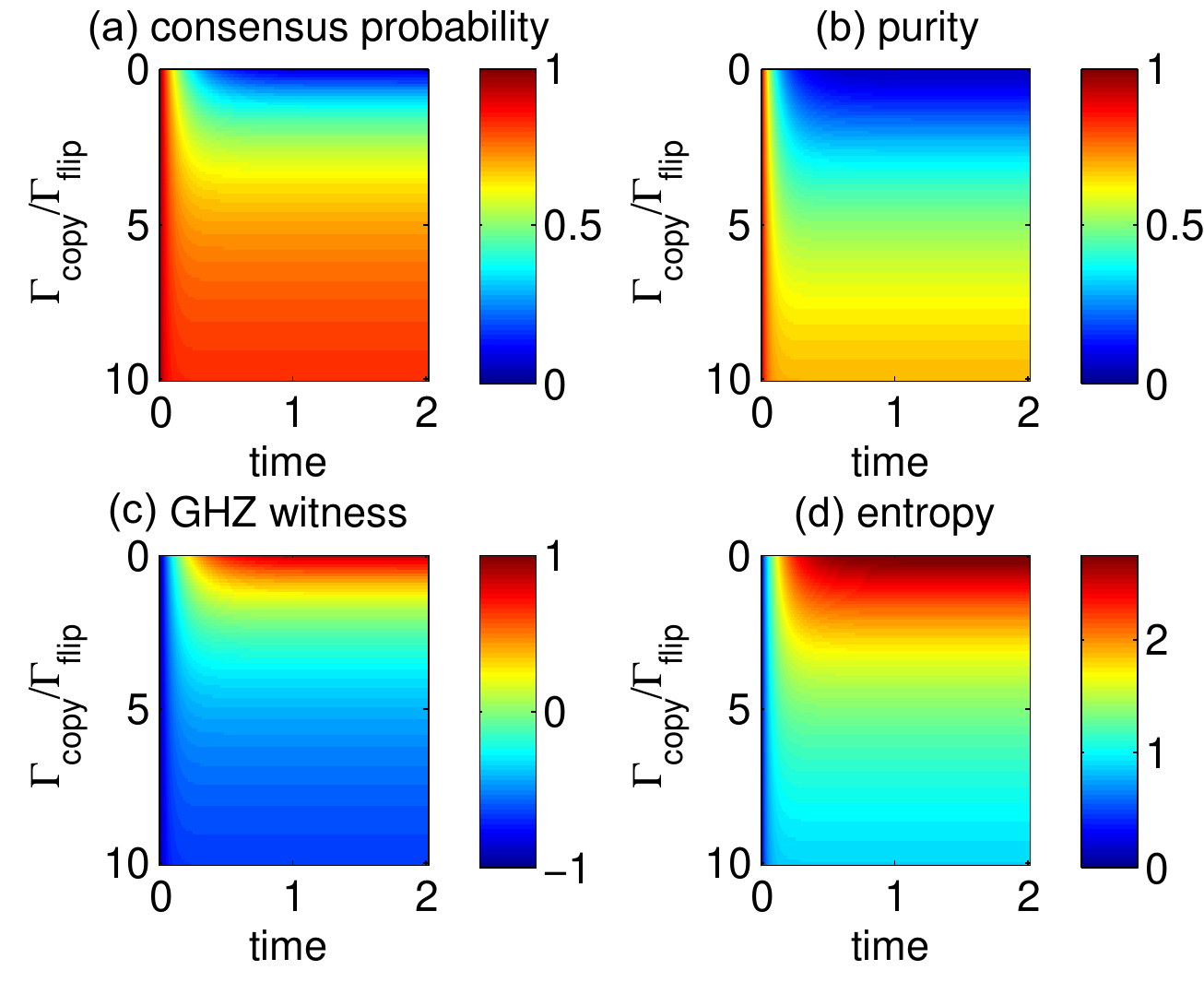}
 \caption{(Color online) Entanglement stabilization against bit-flip noise.
 Here, the initial state is GHZ state, $\ket{00000} +\ket{11111}$.
 When $\Gamma_{\text{copy}} \gg \Gamma_{\text{flip}}$, the consensus (a), entanglement (c), coherence (b), and entropy (d) is protected against the bit-flip noise. The time is measured in $\Gamma_{\text{flip}}^{-1}$.}
 \label{fig:Ent-prot}
 \end{figure}

Fig.~\ref{fig:Ent-prot} shows that the quantum feedback controls protect the GHZ entanglement against the bit-flip noise.
The initial state is now chosen as the 5-qubit GHZ state, $\ket{00000}+\ket{11111}$. (See Appendix~\ref{apx:EntGenSymmInit} for the case of an initial state equal to the symmetric superposition).
 When $\Gamma_{\text{copy}}=0$ the noise destroys, in time, the consensus, the coherence, and the entanglement, while increasing the entropy. When $\Gamma_{\text{copy}}>0$, the effect of the noise is reduced, and the different quantities reach a new stationary value. 
 When $\Gamma_{\text{copy}} \gg \Gamma_{\text{flip}}$, the stationary value of the entanglement is significantly large ($W_{\text{GHZ}}< -0.5$ for $\Gamma_{\text{copy}}/\Gamma_{\text{flip}} >5$).

 We now give an example of a three-qubits case, for an insight of the GHZ stabilization. Let us consider a GHZ state $\rho_{\text{GHZ}}$ which suffered a bit flip with probability $\epsilon$, $(1-\epsilon)\rho_{\text{GHZ}} + \epsilon \rho_{\text{FGHZ}}$. Here $\rho_{\text{FGHZ}}= \sum_i X_i\rho_{\text{GHZ}} X_i/3$. Then, the copy process recovers the flipped state to GHZ state with probability $1/3$; e.g. for the state where qubit 1 is flipped, among 6 possible copy processes only $C_{1\leftarrow 2}$ and $C_{1 \leftarrow 3}$ return the flipped state to the GHZ state. Thus the copy processes contribute to $\dot{\rho}$ by an amount of $\Gamma_{\text{copy}}\epsilon (\rho_{\text{GHZ}}-\rho_{\text{FGHZ}})/3$. The noise converts the GHZ state to the flipped state or the flipped state to the double flipped state. The double flipped state is the same as the GHZ state with probability of $1/3$, or equal to the single-flipped state otherwise. Thus the noise contributes to $\dot{\rho}$ by amount of $(1-\epsilon)\Gamma_{\text{flip}}(\rho_{\text{FGHZ}}-\rho_{\text{GHZ}}) + \epsilon \Gamma_{\text{flip}} [(1/3)\rho_{\text{GHZ}} + (2/3)\rho_{\text{FGHZ}}-\rho_{\text{FGHZ}}]$. A steady state is formed at $\epsilon = 3/(4+\Gamma_\text{copy}/\Gamma_\text{flip})$, satisfying $\dot{\rho}=0$ when summing both contributions.

After showing that the GHZ state can be generated and stabilized against the bit-flip noises, we discuss now how it can be realized in experiments. Let us consider that the qubit states 0 and 1 are realized by the $1/2$-spin up and down state in the $z$ direction, respectively. The initial state of the symmetric superposition $\sum_{s_1,\cdots,s_N=0,1} \ket{s_1 \cdots s_N} = (\ket{0}+\ket{1})^{\otimes N}$ can be prepared by applying a magnetic field in the $x$ direction for a sufficiently large period so that the spins aligns to the $x$ direction. The Zeeman splitting $E_x$ due to the magnetic field should be much larger than the thermal energy broadening $k T$ ($k$ is Boltzmann's constant), as the spin is directed towards the negative $x$ direction with probability $e^{-E_x/k T}$. After the preparation of the initial state, a quantum feedback control $C_{i \leftarrow j}$ operates in a randomly chosen qubit pair (e.g. using a random number generator), in repetitions with a frequency of $\Gamma_{\text{copy}}$.
This quantum feedback control $C_{i\leftarrow j}$ can be performed experimentally, as it requires the C-NOT operation that has been recently implemented between two electron spins by resonantly driving them in a inhomogeneous Zeeman field~\cite{zajac2018resonantly}.
Our system can be realized by applying the scheme to $N+1$ quantum dots, where one quantum dot plays the role of the ancilla.
The regime of the fast copy processes, $\Gamma_{\text{copy}}> \Gamma_{\text{flip}}$, can be reached in the experiments like Ref.~\cite{zajac2018resonantly}, as $\Gamma_{\text{copy}} \sim $ 5 MHz (according to the operation time $\sim$ 200 ns of C-NOT gate) and $\Gamma_{\text{flip}} \sim $0.1 MHz (according to the spin decoherence time of 10 $\mu$s).
As detailed in the Appendix~\ref{apx:EntGenSymmInit},
we find that GHZ-entanglement is also generated when the initial state deviates from the symmetric superposition, e.g. $W_{\text{GHZ}} < -0.5$ when the initial spins are tilted from the $x$ direction by an azimuthal angle less than $0.1\pi$, or when they are directed towards negative $x$ direction by a probability less than $0.1$.

We note that if the consensus dynamics is induced by classical feedback controls, then there is no feature of entanglement generation as the above result, see Appendix~\ref{apx:clsFB}.

\section{Second law of thermodynamics}
\label{sec:2ndLaw}

\subsection{Upper bounds of entropy reduction and work extraction rate}
\label{sec:ubs}

Here we derive the upper bound of the entropy reduction and work extraction rate by a generic class of entangling Maxwell demons.
Our derivations for the bounds are based on the approach of Ref.~\cite{funo2015quantum} and Ref.~\cite{ptaszynski2019thermodynamics}.

We consider a situation in which the working substance, composed of $N$ qubits, is subject to an arbitrary Hamiltonian, arbitrary Markovian dissipations induced by weak coupling to a bath of temperature $T$, and to the action of the entangling Maxwell demon. In the derivation below we use the specific notation of the quantum voter model, but this can be generalized straightforwardly to a generic entangling Maxwell demon.

We first derive the Lindbladian master equation describing the time evolution of the quantum state of the working substance (which will be called ``system'' below). Let $\mathcal{L}_0$ be the Liouvillian in the absence of feedback controls. 
We consider a time step $\Delta t$ which is smaller than the repetition period of the selection and feedback, $\Gamma_{\text{copy}}^{-1}$, and the time scales of $\mathcal{L}_0$, but larger than the operation time of a single quantum feedback control, e.g. the time for completing the three controlled-NOT gates in the case of the quantum voter model.
Then, we obtain the density matrix of the system evolved by both $\mathcal{L}_0$ and the feedback controls,
\begin{equation}
 \label{eq:timeevol-gen}
 \begin{aligned}
 \rho(t+\Delta t) =& \Gamma_{\text{copy}} \Delta t \sum_{i\neq j} C_{i \leftarrow j}(\rho) \\
 &+(1- N_{\text{copy}}\Gamma_{\text{copy}}\Delta t)e^{\mathcal{L}_0 \Delta t}\rho + \mathcal{O}(\Delta t^2).
 \end{aligned}
\end{equation}
Hence, we obtain the master equation,
\begin{equation}
 \label{eq:Lindblad-gen}
 \dot{\rho} = -\frac{i}{\hbar}[H, \rho] + D_{\text{bath}}[\rho] +D_{\text{copy}}[\rho].
\end{equation}
Here $H$ is the system Hamiltonian, $D_{\text{bath}}$ (resp. $D_{\text{copy}}$) is a Lindblad super-operator describing the dissipation of the system induced by its coupling to the bath (resp. multi-particle dissipations due to the feedback control). In the case of the quantum voter model, $D_{\text{copy}}$ is
\begin{equation}
 \label{eq:D-copy}
 D_\text{copy}[\rho] = \Gamma_{\text{copy}} \sum_{i\neq j}
 \Big[ C_{i \leftarrow j}(\rho)-\rho \Big].
\end{equation}

As the dissipations by the bath and the feedback control contribute additively to the master equation, 
the rate of change of system entropy, $\dot{S}$, is the sum of the two contributions. Using $\dot{S}=-\text{Tr}(\dot{\rho}\ln\rho )$, we obtain
\begin{align}
 \label{eq:dS-gen}
 \dot{S} &= \dot{S}_{\text{bath}} + \dot{S}_{\text{copy}} , \\
 \label{eq:dS-bath}
 \dot{S}_{\text{bath}} &\equiv -\text{Tr}(D_{\text{bath}}[\rho] \ln\rho ), \\
 \label{eq:dS-copy}
 \dot{S}_{\text{copy}} &\equiv- \text{Tr}(D_{\text{copy}}[\rho] \ln\rho ). 
\end{align}
Note that $\dot{S}_{\text{bath}}$ (resp. $\dot{S}_{\text{copy}}$) at time $t$ equals the the rate of change of the system entropy assuming that the system $\rho(t)$ is subject only to the dissipation by the bath (resp. feedback controls).

The entropy change rate by the bath is lower-bounded depending on the rate of heat absorption $\dot{Q}$ from the bath as~\cite{ptaszynski2019thermodynamics},
\begin{equation}
 \label{eq:lb-dSdt-bath}
 \dot{S}_{\text{bath}} \ge \dot{Q}/T .
\end{equation}
On the other hand, the entropy reduction rate by the entangling Maxwell demon, $-\dot{S}_{\text{copy}}$, is upper-bounded as 
\begin{equation}
 \label{eq:dSdt-lambda}
 \dot{S}_{\text{copy}} \ge - N_{\text{copy}} \Gamma_{\text{copy}}
 (\overline{I_{\text{QC}}} - \overline{\lambda_{\text{fb}}}),
\end{equation}
(see Appendix~\ref{sec:ubEnt} for its derivation). This is our first main result of this section. 

The upper-bound of the entropy reduction rate, Eq.~(\ref{eq:dSdt-lambda}), is determined by the competition of two quantities.
First, $\overline{I_{\text{QC}}} $ is the quantum-classical mutual information gained in the feedback control averaged for all the possible selections,
\begin{align}
 \overline{I_{\text{QC}}}
 &\equiv N_{\text{copy}} ^{-1}\sum_{i\neq j} I_{\text{QC}}^{(ij)} , \\
 I_{\text{QC}}^{(ij)}
 &= S(\rho) - \sum_{k=0,1}p_k^{(ij)} S(\rho_k^{(ij)}) 
\end{align}
where $I_{\text{QC}}^{(ij)}$ is the mutual information given that the qubits $i$ and $j$ are selected.
Second, $\overline{\lambda_{\text{fb}}}$ is the absolute irreversibility of a stochastic feedback control (applied to a random pair of qubits), see Sec.~\ref{sec:lambda}.

Using Eqs.~(\ref{eq:lb-dSdt-bath}) and (\ref{eq:dSdt-lambda}), we obtain 
\begin{equation}
 \label{eq:dS-gen-lb}
 \dot{S} \ge \frac{\dot{Q}}{T} + N_{\text{copy}} \Gamma_{\text{copy}} (-\overline{I_{\text{QC}}} +\overline{\lambda}_{\text{fb}}).
\end{equation}
Replacing $\dot{Q} = \dot{E}-\dot{W}$, as given by the first law of thermodynamics ($\dot{E}$ is energy change rate of the system), into Eq.~(\ref{eq:dS-gen-lb}), we obtain the upper bound of the work extraction rate $-\dot{W}$,
\begin{equation}
 \label{eq:2ndLaw-W}
 \dot{W}-\dot{F} \ge k T N_{\text{copy}}\Gamma_{\text{copy}}
 (-\overline{I_{\text{QC}}} +\overline{\lambda_{\text{fb}}}),
\end{equation}
This is our second main result of this section.
$\dot{F} = \dot{E} -T \dot{S}$ is the rate of change of nonequilibrium free energy~\cite{parrondo2015thermodynamics}.

\subsection{Absolute irreversibility}
\label{sec:lambda}

Here we show that the absolute irreversibility $\overline{\lambda_{\text{fb}}}$ of a stochastic feedback control is related to how much the feedback-operated states are different for distinct selection.

As we mentioned earlier, the absolute irreversibility is defined as the sum of probabilities of the backward trajectories which do not have a corresponding forward trajectory~\cite{funo2015quantum}.
The initial state of the forward trajectory is the nonequilibrium state $\rho$ of the system at the time for evaluating the 
entropy reduction (or work extraction) rate.
The final state of the forward trajectory is the result of a stochastic feedback control without knowing which particles were selected, 
\begin{equation}
 \label{eq:C-rho}
 C(\rho) \equiv \frac{1}{N_{\text{copy}} } \sum_{i\neq j} C_{i \leftarrow j}(\rho).
\end{equation}
This state equals the initial state $C(\rho)=\rho_r$ of the backward trajectory, the reference state in the language of the fluctuation theorem (see Appendix~\ref{sec:ubEnt}).

The absolute irreversibility $\overline{\lambda_{\text{fb}}}$ of a stochastic feedback control is the average of the absolute irreversibility given that it is known which particles were selected,
\begin{align}
 \label{eq:lambda-av}
 \overline{\lambda_{\text{fb}}}
 &= \frac{1}{N_{\text{copy}}}
 \sum_{i \neq j} \lambda_{\text{fb}}^{(ij)}, \\
 \label{eq:lambda-ij}
 \lambda_{\text{fb}}^{(ij)}
 &\equiv \sum_{k=0,1} p_k^{(ij)} \text{Tr} [\Pi_{\text{null}(\rho_k^{(ij)})} U_k^{(i)\dagger} \rho_r U_k^{(i)}] .
\end{align}
$\lambda_{\text{fb}}^{(ij)}$ is the absolute irreversibility given that the selected particles were $i$ and $j$, and 
 $\Pi_{\text{null}(\rho_k^{(ij)})}$ is the projection operator onto the null space of the post-measurement state $\rho_k^{(ij)}$.
Using Eqs.~(\ref{eq:C-rho})--~(\ref{eq:lambda-ij}) and the relation
\begin{equation}
 \label{eq:P-sup-U}
 U_k^{(i)} \Pi_{\text{null}(\rho_k^{(ij)})} U_k^{(i)\dagger}
 = \Pi_{\text{null}(U_k^{(i)} \rho_k^{(ij)}U_k^{(i)\dagger})},
\end{equation}
we obtain the absolute irreversibility for a stochastic feedback control,
\begin{equation}
 \label{eq:lambda-C}
 \overline{\lambda_{\text{fb}}}
 = \sum_{i\neq j, l\neq m} \frac{1}{N_{\text{copy}}^2}\sum_{k,q=0,1}
 p_k^{(ij)} p_q^{(lm)} \text{Tr} \big[\Pi_{\text{null}(\rho_{\text{fin},k}^{(ij)})}
 \rho_{\text{fin},q}^{(lm)} \big] .
\end{equation}
$\rho_{\text{fin},k}^{(ij)} = U_k^{(i)} \rho_k^{(ij)}U_k^{(i)\dagger}$ is the feedback-operated state given that the selected particles were $i$ and $j$, and measurement outcome was $k$.

Eq.~(\ref{eq:lambda-C}) shows that that the absolute irreversibility $ \overline{\lambda_{\text{fb}}}$ is related to how much the feedback-operated states are different for distinct selection.
The term $ \text{Tr} [\Pi_{\text{null}(\rho_{\text{fin},k}^{(ij)})}
 \rho_{\text{fin},q}^{(lm)} ] $ quantifies how much the result of two different feedback controls are different;
it is $1$ when the support of the two feedback-operated states are orthogonal, and $0$ when identical.

We also find that, the absolute irreversibility for a pure state $\rho$ is determined by how much the state decoheres by the stochastic feedback control,
\begin{equation}
 \label{eq:lambda-pure}
 \overline{\lambda_\text{fb}}
 =1 - \text{Tr} \left[ C(\rho)^2 \right].
\end{equation}
This is proved as follows.  From Eq.~(\ref{eq:lambda-C}), we use $\Pi_{\text{null}(\rho^{(ij)}_{\text{fin},k})}= \mathbb{1}- \Pi_{\text{sup}(\rho^{(ij)}_{\text{fin},k})}$,  where $\Pi_{\text{sup}(\rho^{(ij)}_{\text{fin},k})} $ is the projection operator onto the support of $\rho^{(ij)}_{\text{fin},k}$.
As $\rho$ is a pure state, $\rho_{\text{fin},k}^{(ij)}$ is also a pure state, and we simplify $\Pi_{\text{sup}(\rho_{\text{fin},k}^{(ij)})} = \rho_{\text{fin},k}^{(ij)}$. Then, after summing $k,q$ and using that $ \sum_k p_k^{(ij)}\rho_{\text{fin},k}^{(ij)}=C_{i \leftarrow j}(\rho)$, $\sum_q p_q^{(lm)}\rho_{\text{fin},q}^{(lm)}=C_{l \leftarrow m}(\rho)$, we obtain
\begin{align}
 \overline{\lambda_{\text{fb}}}
 &= \frac{1}{N_{\text{copy}}^2}\sum_{i\neq j,l\neq m} 
 \Big[1 - \text{Tr} \big(C_{i \leftarrow j}(\rho) C_{l \leftarrow m}(\rho)\big)\Big], \label{eq:lambda-pure-S2} \\
 &= 1 - \text{Tr} \Big[ \frac{1}{N_{\text{copy}}^2}\sum_{i\neq j,l\neq m} C_{i \leftarrow j}(\rho) C_{l \leftarrow m}(\rho) \Big]. \label{eq:lambda-pure-S3}
\end{align}
The final equality is equivalent to Eq.~(\ref{eq:lambda-pure}) after summing $i,j, l,m$ and using Eq.~(\ref{eq:C-rho}).

\subsection{Competition between $\overline{I_{\text{QC}}}$ and $\overline{\lambda_{\text{fb}}}$ }
\label{sec:comp}

\begin{figure}[t]
 \centering \includegraphics[width=\columnwidth]{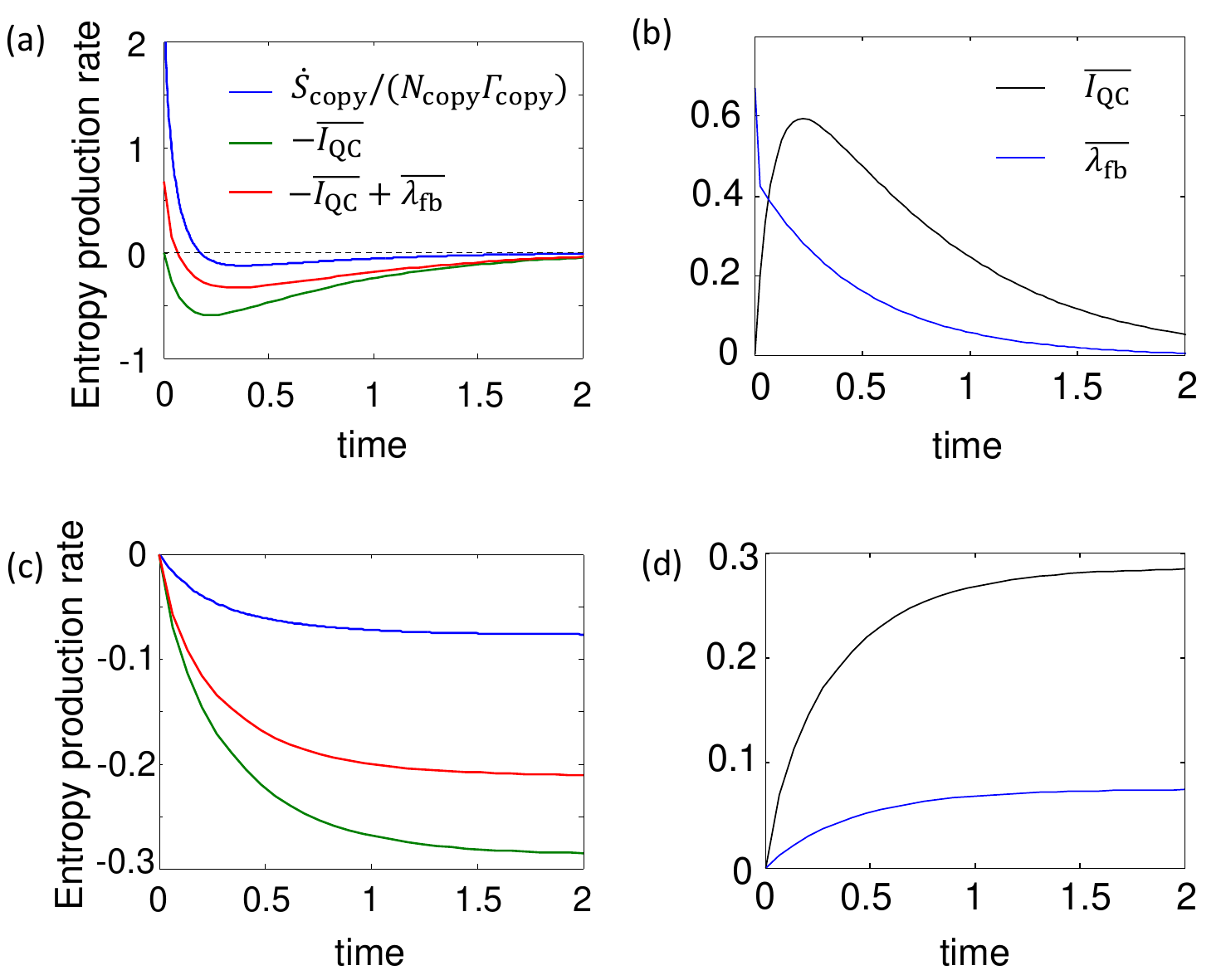}
 \caption{(Color online) Competition of the quantum-classical mutual information $\overline{I_{\text{QC}}}$ and the absolute irreversibility $\overline{\lambda_{\text{fb}}}$. 
 (a) Entropy production rate by copy processes in $\Gamma_{\text{flip}}=0$ and a symmetric initial state, the situation of Fig.~2.
 (b) Competition of $\overline{I_{\text{QC}}}$ and $\overline{\lambda_{\text{fb}}}$.
 (c)-(d) Plots corresponding to (a)-(b) in $\Gamma_{\text{flip}}= 0.1 \Gamma_{\text{copy}}$ and a GHZ initial state, the situation of Fig.~3.
 The time is measured in $\Gamma_{\text{copy}}^{-1}$.
 }
 \label{fig:S-I}
 \end{figure}
 
The result of the quantum voter model can be understood in terms of the competition between $\overline{I_{\text{QC}}}$ and $\overline{\lambda_{\text{fb}}}$.
Figure~\ref{fig:S-I} (a)--(b) shows the entropy production rate (i.e. the negative value of the reduction rate), in the absence of bit-flip noises and a symmetric initial state, the situation of Fig.~\ref{fig:ghzN}.
Around the initial time, it is $\overline{\lambda_{\text{fb}}}>0$ because different choices of copy pairs result in different feedback-operated states. For example, at the initial time, $C_{1\leftarrow 2}(\rho)= \ket{\Phi_{+}} \otimes \ket{+++}$ is different from
 $C_{4\leftarrow 5}(\rho)=\ket{+++}\otimes \ket{\Phi_{+}}$, as their overlap is $1/4$,
 where
 \begin{align}
 \ket{\pm} &\equiv \ket{0}\pm \ket{1} , \\
 \ket{\Phi_+} &\equiv \ket{00}+\ket{11} = \ket{++}+\ket{--} ,
 \end{align}
$\ket{\Phi_+}$ being the maximally entangled two-qubit state.
 It can be proved (see Appendix \ref{apx:lambda-symm} for the derivation) that for the symmetric state one finds $\overline{\lambda_{\text{fb}}}= 3/4\times [1-2/N_\text{copy}]$, taking the value $\overline{\lambda_{\text{fb}}}=0$ at $N=2$ and monotonically increasing up to $3/4$ as $N$ increases.
 Meanwhile, initially $\overline{I_{\text{QC}}}=0$ because for any $C_{i \leftarrow j}$, there is no decrease of entropy when the measurement outcomes are known;
 $S(\rho)=0$ and $S(\rho_k^{(ij)})=0$ as the initial state $\rho$ is a pure state. Therefore, the absolute irreversibility dominates over the quantum-classical mutual information and the entropy increases by the copy processes. In addition, the purity decreases [see Fig.~\ref{fig:ghzN} (b)] due to the non-vanishing $\overline{\lambda_{\text{fb}}}$, as predicted by Eq.~(\ref{eq:lambda-pure}). 
 As time increases, the system gets closer to the consensus state [see Fig.~\ref{fig:ghzN} (a)].
 Then, most copy processes $C_{i \leftarrow j}$ do not change the state, hence $\overline{\lambda_{\text{fb}}} \rightarrow 0$.
 When $\overline{\lambda_{\text{fb}}} < \overline{I_{\text{QC}}}$, the entropy is reduced by the copy processes.

 Figs.~\ref{fig:S-I} (c)-(d) show the entropy production rate in the presence of bit-flip noise and a GHZ initial state, the situation of Fig.~\ref{fig:Ent-prot}. In this case, the initial state yields $\overline{\lambda_{\text{fb}} }= 0$, as the state is in the perfect consensus and
 $\overline{I_{\text{QC}}}=0$ as the initial state is a pure state. As time increases, the noises induce lack of consensus and decoherence.
 Therefore both $\overline{\lambda_{\text{fb}} }$ and $\overline{I_{\text{QC}}}$ increase. In this case,
 $\overline{I_{\text{QC}}}> \overline{\lambda_{\text{fb}} }$ at all times and the entropy is always reduced by the copy processes.
Note that the inequality Eq.~(\ref{eq:dSdt-lambda}) is always satisfied.
When ignoring the absolute irreversibility $\overline{\lambda_{\text{fb}}}$, the lower bound determined by the quantum-classical mutual information predicts the entropy reduction in a too much optimistic way; see the green curve in panel \ref{fig:S-I}.(a).

\subsection{Example of the work extraction by the entangling Maxwell Demon}
\label{sec:egDemon}

Finally, we provide a simple example showing that the entangling Maxwell demon can indeed convert heat from the thermal bath to work using the information $\overline{I_{\text{QC}}}$.

Consider two qubits interacting by a Hamiltonian
\begin{equation}
 \label{eq:H-int}
 H=-J Z_1 \otimes Z_2,
\end{equation}
and weakly coupled to a bath of temperature $T$ which induces the bit-flip noise.
Here $Z_i$ is the Pauli matrix measuring the spin of qubit $i$ in the z-direction, and $J (>0)$ is the interaction strength between the two qubits.
The bit-flip noise can be realized when a) each qubit observable $X_i$ is weakly coupled to the bosonic bath~\cite{schaller2014open}, and b) the interaction between the qubits is weak $J \ll \hbar \Gamma_{\text{flip}} $~\cite{cattaneo2019local}. Note that the second condition is necessary to ensure that the bath induces independent local dissipations for each qubit. The time evolution of the qubits is governed by Eq.~(\ref{eq:rho_C_N}) with additional term $-(i/\hbar)[H, \rho]$,
\begin{equation}
 \label{eq:L-2qbs}
\begin{aligned}
 \dot{\rho} = &- \frac{i}{\hbar} [H,\rho] \\
 &+ \Gamma_{\text{copy}}\sum_{i\neq j} (C_{i \leftarrow j} (\rho)-\rho)
 +\Gamma_{\text{flip}} \sum_i (X_i \rho X_i -\rho).
\end{aligned}
\end{equation}
Once the two qubits are driven into the steady state near the maximally-entangled consensus state
\begin{equation}
 \label{eq:rhoC}
\rho_C \equiv \frac{1}{2} \big(\ket{00}+\ket{11}\big)\big(\bra{00}+\bra{11}\big)
\end{equation}
 by the fast copy processes, they convert heat absorbed from the bath to work; the flip noise accompanies a heat absorption of $2J$ and the copy process extracts work of the same amount $2J$ into driving field which operates the measurement and feedback operation (i.e. three controlled-NOT gates)~\cite{elouard2017extracting}.

Now, we calculate the work extraction rate $\dot{W}_\text{ext}\equiv -\dot{W}$ for the steady state $\rho_{\text{st}}$ formed near the maximally entangled state.
The work extraction rate is equal to the heat absorption rate due to the first law of thermodynamics,
\begin{equation}
 \label{eq:dW-2qb-Q}
 \dot{W}_{\text{ext}} = \Gamma_{\text{flip}}
 \sum_i \Big[ \text{Tr} (X_i \rho_\text{st} X_i H)
 -\text{Tr} (\rho_{\text{st}}H ) \Big].
\end{equation}
We find that the steady state satisfies
\begin{align}
 \label{eq:rho-st}
 \rho_{\text{st}}&=(1-\epsilon)\rho_{\text{C}} + \epsilon\rho_{\text{FC}} , \\
 \rho_{\text{FC}} &= \frac{1}{2} \big(\ket{01}+\ket{10}\big)\big(\bra{01}+\bra{10}\big) , \\
 \label{eq:epsilon} 
 \epsilon &= \frac{\Gamma_{\text{flip}}}{\Gamma_{\text{copy}}+2\Gamma_{\text{flip}}},
\end{align}
where $\rho_{\text{FC}}$ is one-bit-flipped $\rho_{\text{C}}$, see Appendix \ref{apx:SSheatEngine} for the derivation.
After simple algebra using Eq.~(\ref{eq:XrX}), $\text{Tr}(\rho_\text{C} H) = -J$, and $\text{Tr}(\rho_\text{FC} H) = J$, we obtain the work extraction rate
\begin{equation}
 \label{eq:dW-2qb}
 \dot{W}_{\text{ext}} = 2J \Gamma_{\text{flip}} (1- 2 \epsilon).
\end{equation}

We compare the work extraction rate, Eq.~(\ref{eq:dW-2qb}), with the upper bound $\dot{W}_{\text{ext,ub}}=2kT \Gamma_{\text{copy}} (\overline{I_{\text{QC}}} -\overline{\lambda_\text{fb}})$ dictated by the second law, Eq.~(\ref{eq:2ndLaw-W}).
The information gain $\overline{I_{\text{QC}}} =-\epsilon\ln \epsilon -(1-\epsilon)\ln(1-\epsilon)\equiv h(\epsilon)$ is the binary Shannon entropy in nats, because the initial state $\rho_{\text{st}}$ has an entropy equal to $h(\epsilon)$ and 
the post-measurement states are pure states with vanishing entropy. 
The absolute irreversibility $\overline{\lambda_\text{fb}}$ vanishes according to Eq.~(\ref{eq:lambda-C});
The feedback-operated states $\rho_{\text{fin},k}^{(ij)}$ are all equal to $\rho_\text{C}$ and
$ \text{Tr} \big[\Pi_{\text{null}(\rho_{\text{fin},k}^{(ij)})}
 \rho_{\text{fin},q}^{(lm)} \big]=0$ for any $i \neq j$ and $k,q \in [0,1]$.
 Therefore, we obtain the upper bound dictated by the second law,
 \begin{equation}
 \label{eq:dW-2qb-ub}
 \dot{W}_{\text{ext,ub}} = 2 kT \Gamma_{\text{copy}} h(\epsilon).
 \end{equation}

 The second law, $\dot{W}_{\text{ext}} \le \dot{W}_{\text{ext,ub}}$, is verified when examining the prerequisites of the above results. Using $\Gamma_{\text{flip}}/\Gamma_{\text{copy}}= \epsilon/(1-2\epsilon)$, the relevant ratio becomes
\begin{equation}
 \frac{\dot{W}_{\text{ext}}}{\dot{W}_{\text{ext,ub}}}
 = \frac{J}{kT} \frac{\epsilon}{h(\epsilon)}.
\end{equation}
The factor $\epsilon/h(\epsilon)$ is smaller than 1 in the allowed range of $\epsilon \in [0, 0.5] $.
We examine the other factor by further factorizing, $J/kT= (J/\hbar\Gamma_{\text{flip}})(\hbar\Gamma_{\text{flip}}/kT) $. The first factor $J/\hbar\Gamma_{\text{flip}}$ is much smaller than 1 due to the condition that the bath induces the independent local dissipations.
The second factor $\hbar\Gamma_{\text{flip}}/kT$ is also much smaller than 1 as $\hbar \Gamma_{\text{flip}}= \mathcal{O}[\lim_{\omega\rightarrow 0} G(\omega) (n_B(\omega) +1)] = G'(0) kT $ \cite{schaller2014open}, where $G(\omega)$ is the spectral coupling density characterizing the qubit-bath coupling and $G'(0) \ll 1$ due to the weak coupling condition.

\section{Conclusion}
\label{sec:conclusion}

We have introduced a generic class of Maxwell demon which generates and stabilizes a many-body entanglement in the working substance.
To understand the quantum dynamics, we introduced a quantum version of the classical noisy voter model used in the context of opinion dynamics. 
As a result, the GHZ entanglement in the working substance was generated and stabilized against the bit-flip noises.
A non-monotonic behavior of the time-evolution of the purity and entropy could be understood in terms of a competition between the information gain and the absolute irreversibility of the feedback control.
We discussed how the quantum voter model can be realized in the semiconductor quantum dots and AC voltage-driven gates.

We have also formulated the second law of thermodynamics under the presence of the entangling Maxwell demon. We have obtained that upper bounds of the entropy reduction and the work extraction rates are determined by the competition between the information gain and
the absolute irreversibility of the feedback control. Our finding for the upper bound for the entropy reduction rate will be helpful for stabilizing a many-body entangled state. The upper bound for the work extraction will be valuable for the exploration of the quantum information engine whose working substance is in a many-body entangled state.

 As a final remark, we compare our GHZ-entanglement stabilization protocol based on the two-particle dissipations to others in the steady-state engineering. 
In the quantum optics community, another type of steady-state engineering based on irreversible population transfer through optical pumping is studied in the trapped ions and Rydberg atom setups~\cite{wiseman2009quantum,zhang2017quantum,cho2011optical,lin2013dissipative,reiter2016scalable}, optionally aided by continuous feedback control~\cite{stevenson2011engineering,feng2011generating,morigi2015dissipative}.
In comparison to this, our protocol has the merit of being broadly applicable to any quantum system (not necessarily trapped ions or Rydberg atom setup) with the only requirement of the possibility of a controlled-NOT gate operation, a most basic operation implementable in diverse experimental setups.
On the other hand, continuous error corrections~\cite{paz1998continuous,ahn2002continuous,sarovar2004practical} also have been proposed to stabilize an arbitrary state against decoherence.
In comparison to this, our protocol can be more useful in a system where the number of qubits is limited, as it only requires one additional ancilla qubit for the quantum feedback control.

{\acknowledgements} We thank Gonzalo Manzano, Roberta Zambrini, and Gian Luca Giorgi for useful discussions and comments.
 Partial financial support has been received from the Agencia Estatal de Investigaci\'on (AEI, MCI, Spain) and Fondo Europeo de Desarrollo Regional (FEDER, UE), under Project PACSS (RTI2018-093732-B-C21/C22) and the Maria de Maeztu Program for units of Excellence in R\&D (MDM-2017-0711).
 \newpage

\appendix

\section{Nearest-neighbor connectivity in the quantum voter model}
\label{apx:1Dnn}
Here we present how the entanglement generation in the quantum voter model changes when the copy pair selection occurs in the one-dimensional  nearest-neighbor connectivity instead of the all-to-all scenario considered in the main text.
In this connectivity, a copy process $C_{i \leftarrow j}$ of randomly chosen pair $i$ and $j$ of nearest neighbor is induced with rate $\Gamma_{\text{copy}}$. We consider periodic boundary conditions, hence the allowed choices for $j$ are $j = i \pm 1$ for $i \in [2, \cdots N-1]$, $j=2,N$ for $i=1$, and $j=1, N-1$ for $i=N$.

\begin{figure}[h]
 \centering
\includegraphics[width=0.9\columnwidth]{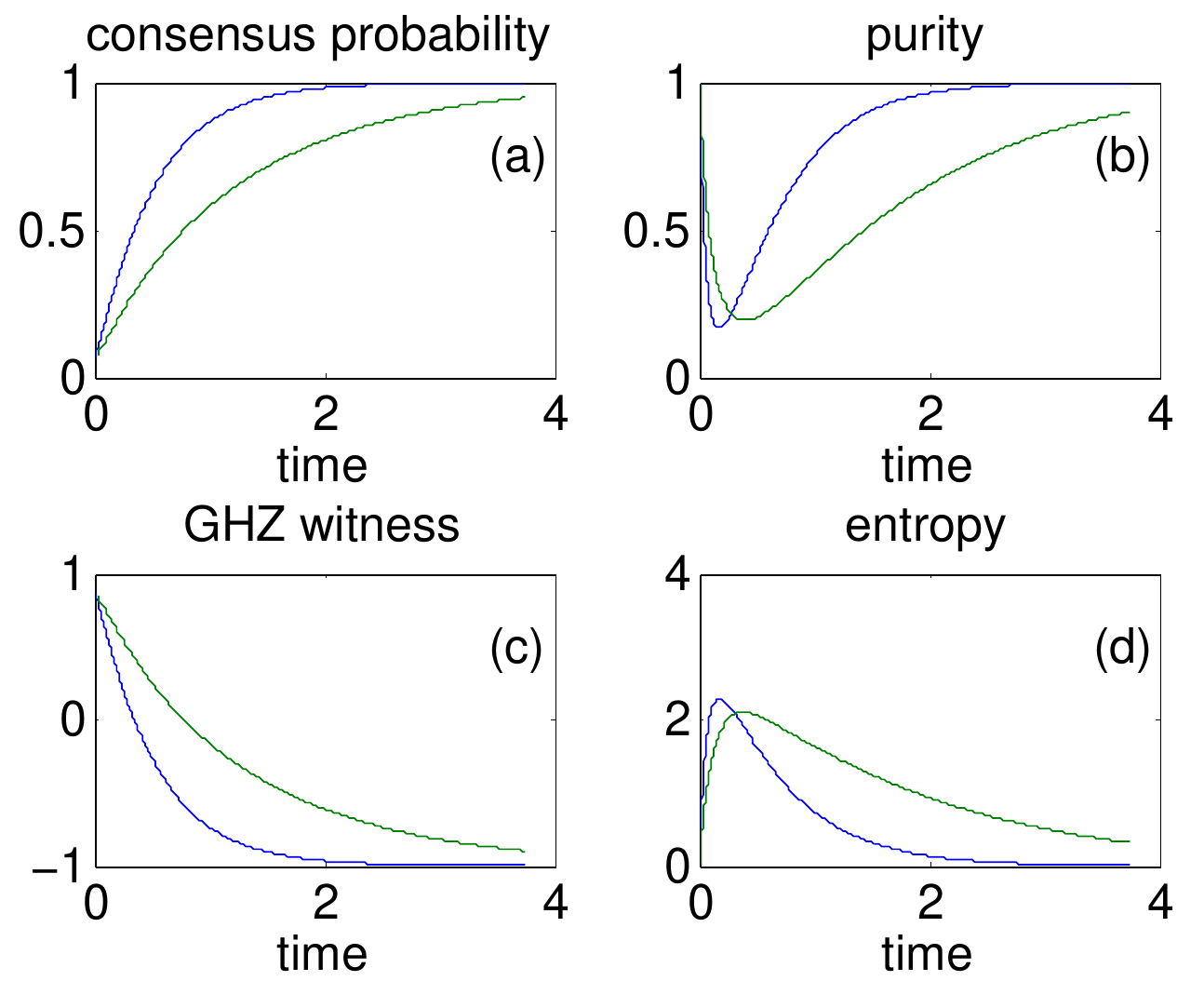}
\caption{(Color online) Comparison of the GHZ entanglement generation in the all-to-all (blue curves) and one-dimensional (green curves) connectivity.
Here, $\Gamma_{\text{flip}}=0$ and the initial state is the symmetric superposition, the situation of Fig.~\ref{fig:ghzN}. The time is measured in $\Gamma_{\text{copy}}^{-1}$.}
 \label{fig:ATA-ring}
\end{figure}

Fig.~\ref{fig:ATA-ring} shows the comparison of the GHZ entanglement generation for the two considered connectivities, in the absence of noise and the symmetric initial state, the situation of Fig.~\ref{fig:ghzN}.
The result shows that the all-to-all connectivity is more efficient for generating the GHZ entanglement, as the GHZ witness (c) of the all-to-all connectivity approaches to $-1$ faster.

\section{Derivation of Eq.~(\ref{eq:rho_C_N}) and its Lindblad form}
\label{apx:Lindblad-X}
 
Here we show the derivation of Eq.~(\ref{eq:rho_C_N}) and that it is in the Lindblad form.

We consider the evolution of the density matrix $\rho(t)$ during a time step $\Delta t$ which is smaller than $\Gamma_{\text{copy}}$ and $\Gamma_{\text{flip}}$ but larger than the operation time of a single copy process, i.e. the time for completing three controlled-NOT gates in Fig.~\ref{fig:setup}(b). During the time step $\Delta t$, the copy process $C_{i \leftarrow j}$ occurs with probability $\Gamma_{\text{copy}} \Delta t$, the flip process $X_i$ occurs with probability $\Gamma_{\text{flip}}\Delta t$, and the state remains the same otherwise. Hence the time-evolved state is
\begin{equation}
 \label{eq:rho-dt}
 \begin{aligned}
 \rho(t+\Delta t) =&\Gamma_{\text{copy}}\Delta t \sum_{i\neq j}
 C_{i \leftarrow j}(\rho) + \Gamma_{\text{flip}} \Delta t \sum_i X_i \rho X_i \\
 & +( 1- N_{\text{copy}} \Gamma_{\text{copy}} - N \Gamma_{\text{flip}}) \rho.
 \end{aligned}
\end{equation}
Dividing this equation by $\Delta t$ and collecting $\dot{\rho}= \dfrac{\rho(t+\Delta t)-\rho(t)}{\Delta t}$, we obtain Eq.~(\ref{eq:rho_C_N}). That equation is in the Lindblad form,
\begin{equation}
 \label{eq:ME-lindblad}
 \begin{aligned}
 \dot{\rho} =& \Gamma_{\text{copy}} \sum_{i\neq j} \sum_{k=0,1} \Big(L^{(ij)}_k \rho L_k^{(ij)\dagger} -\frac{1}{2} \Big\{ \rho, L_k^{(ij)\dagger} L_k^{(ij)} \Big\} \Big) \\
 & + \Gamma_{\text{flip}} \sum_i \Big( L_i \rho L_i^\dagger
 -\frac{1}{2} \Big\{ \rho, L_i^\dagger L_i \Big\} \Big)
 \end{aligned}
\end{equation}
where the Lindblad operators for the copy and flip processes are $L_k^{(ij)}= U_k^{(i)} M_k^{(ij)}$ and $L_i = X_i$, respectively. $\{\cdots, \cdots\}$ is the anticommutator. This form is equivalent to Eq.~(\ref{eq:rho_C_N}) because $\sum_k L_k^{(ij)\dagger}L_k^{(ij)} = \sum_k M_k^{(ij)} U_k^{(i)\dagger}U_k^{(i)} M_k^{(ij)}= \sum_k M_k^{(ij)}=\mathbb{1} $, $L_i^\dagger L_i = X_i^2 = \mathbb{1}$, and
\begin{align}
 \sum_k \frac{1}{2} \Big\{\rho, L_k^{(ij)\dagger}L_k^{(ij)}\Big\}
 &= \frac{1}{2} \{\rho, \mathbb{1}\} = \rho, \\
 \frac{1}{2} \{\rho, L_i^\dagger L_i\} &= \frac{1}{2} \{ \rho, \mathbb{1} \} = \rho.
\end{align}

\section{Quantum voter model with presence of noise and symmetric the initial state}
\label{apx:EntGenSymmInit}

Here, we present a supplemental result in addition to Fig.~\ref{fig:ghzN} and \ref{fig:Ent-prot}, the dynamics of the quantum voter model in the presence of bit-flip noises for the initial state of the symmetric superposition, $ \sum_{s_1,\cdots s_N=0,1}\ket{s_1\cdots s_N}$.

\begin{figure}[h]
 \centering \includegraphics[width=0.95\columnwidth]{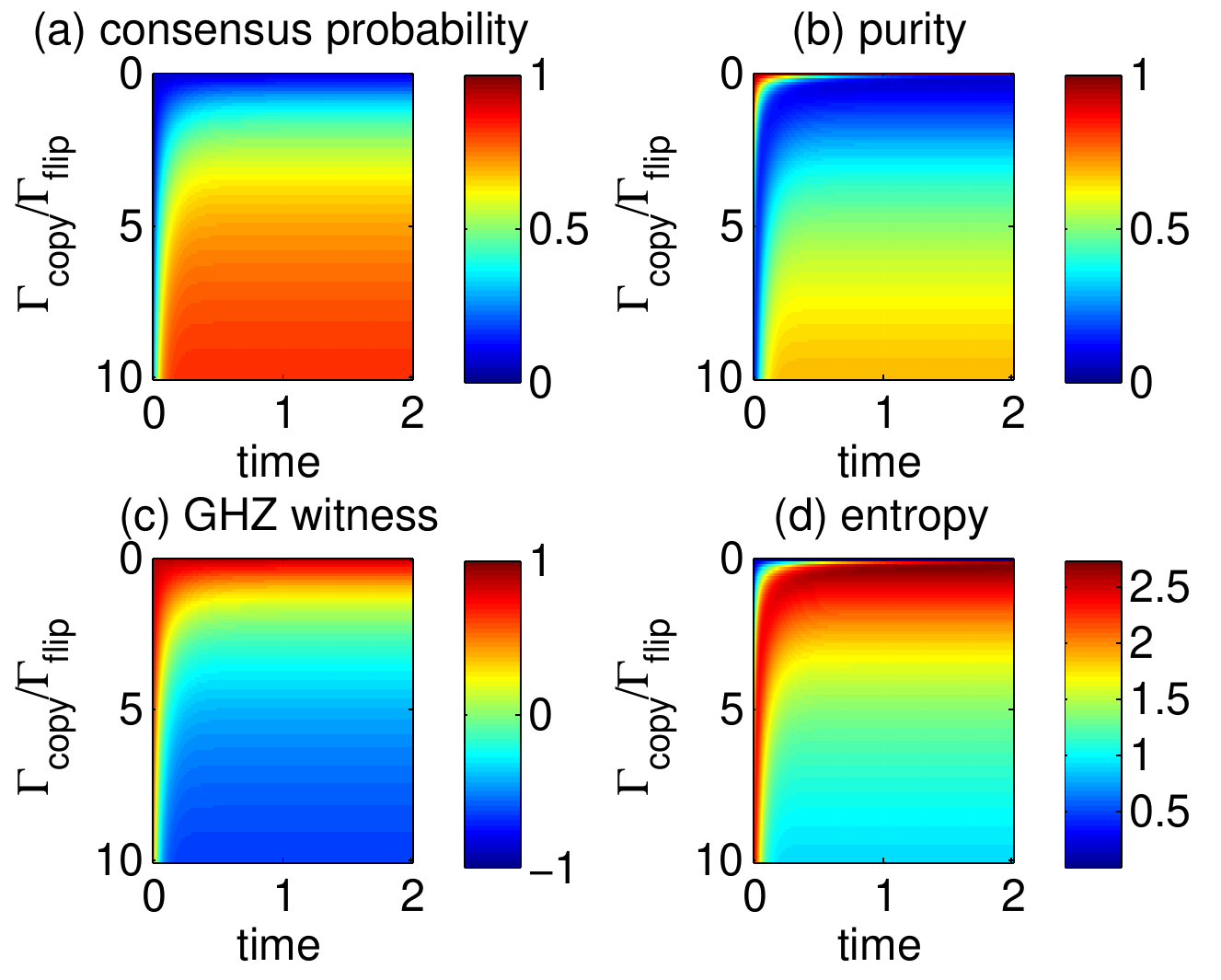}
 \caption{(Color online) GHZ entanglement generation and stabilization in the quantum voter model with the presence of noise and the symmetric initial state. The time is measured in $\Gamma_{\text{flip}}^{-1}$.} 
 \label{fig:ent-gen-tf} 
\end{figure}

Fig.~\ref{fig:ent-gen-tf} shows that GHZ entanglement is generated and stabilized for this case
when $\Gamma_{\text{copy}}\gg \Gamma_{\text{flip}}$, namely $W_{\text{GHZ}}<-0.5$ is achieved at the stationary value when $\Gamma_{\text{copy}}>10 \Gamma_{\text{flip}}$.

In addition, we show that GHZ entanglement is generated and stabilized even when the initial state is deviated from the symmetric superposition. 
We consider two types of deviations which are relevant in the experimental preparation of the symmetric state using an external magnetic field in the $x$ direction (see the main text).
First, the magnetic field can be tilted from the $x$ direction by a azimuthal angle $\theta$.
Then, the initial (pure) state of $ [ (\ket{+} + e^{i \theta/2}\ket{-})]^{\otimes N} $ is prepared. 
Second, a spin can be directed towards the negative $x$ direction due to thermal fluctuations with probability $p_{x,\text{down}}\equiv e^{-E_x/k T}$.
Then, the initial (mixed) state of $[(1-p_{x,\text{down}})\ket{+}\bra{+} +p_{x,\text{down}}\ket{-}\bra{-}]^{\otimes N}$ is prepared.

\begin{figure}[h]
 \centering \includegraphics[width=0.95\columnwidth]{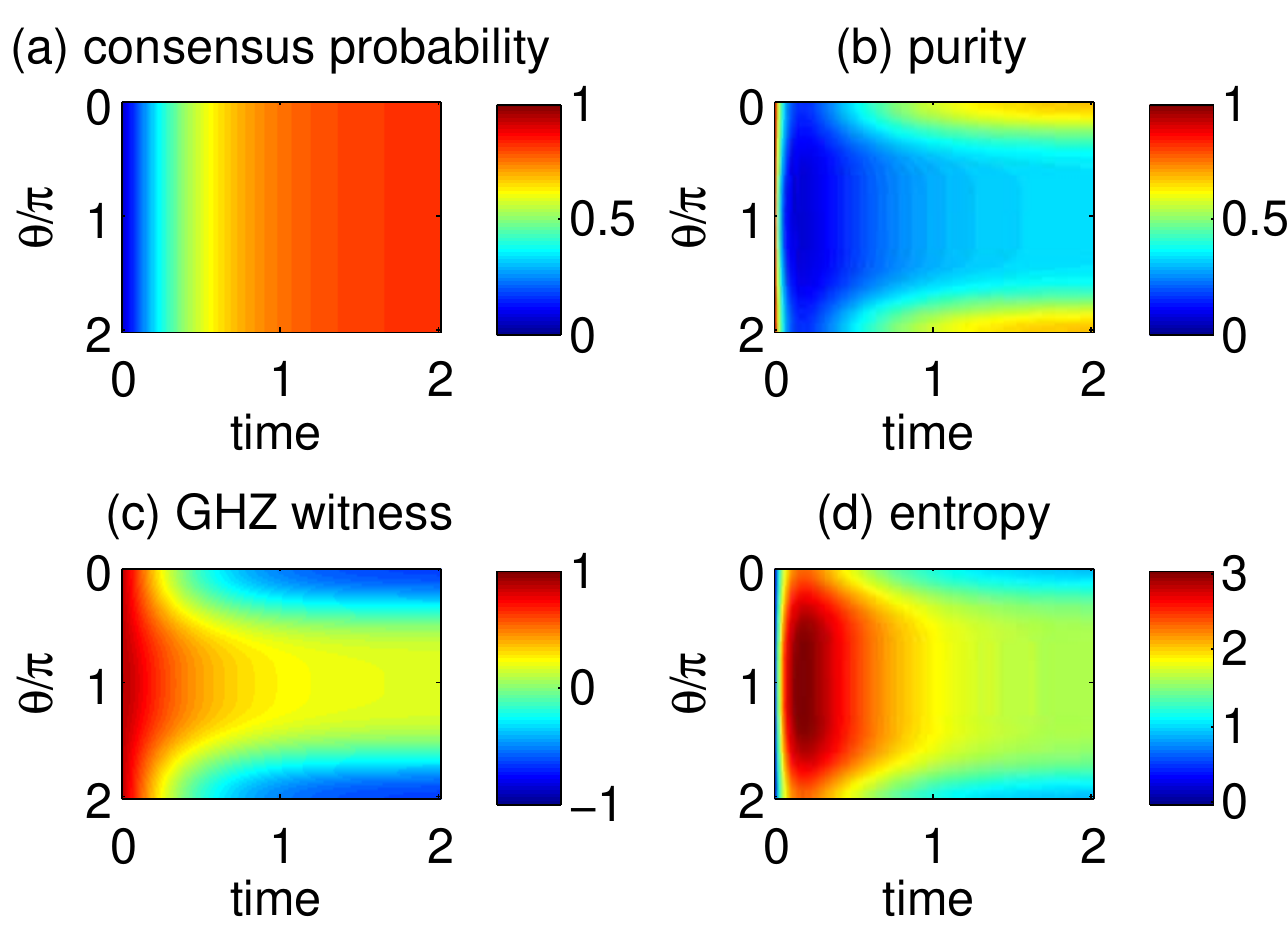}
 \caption{(Color online) Quantum consensus dynamics for the initial state
$ [ (\ket{+} + e^{i \theta/2}\ket{-})/\sqrt{2}]^{\otimes N} $.
Time is measured in $\Gamma_{\text{copy}}^{-1}$. Here, $\Gamma_{\text{flip}}=0.1\Gamma_{\text{copy}}$.}
 \label{fig:init-theta}
\end{figure}

\begin{figure}[h]
 \centering \includegraphics[width=0.95\columnwidth]{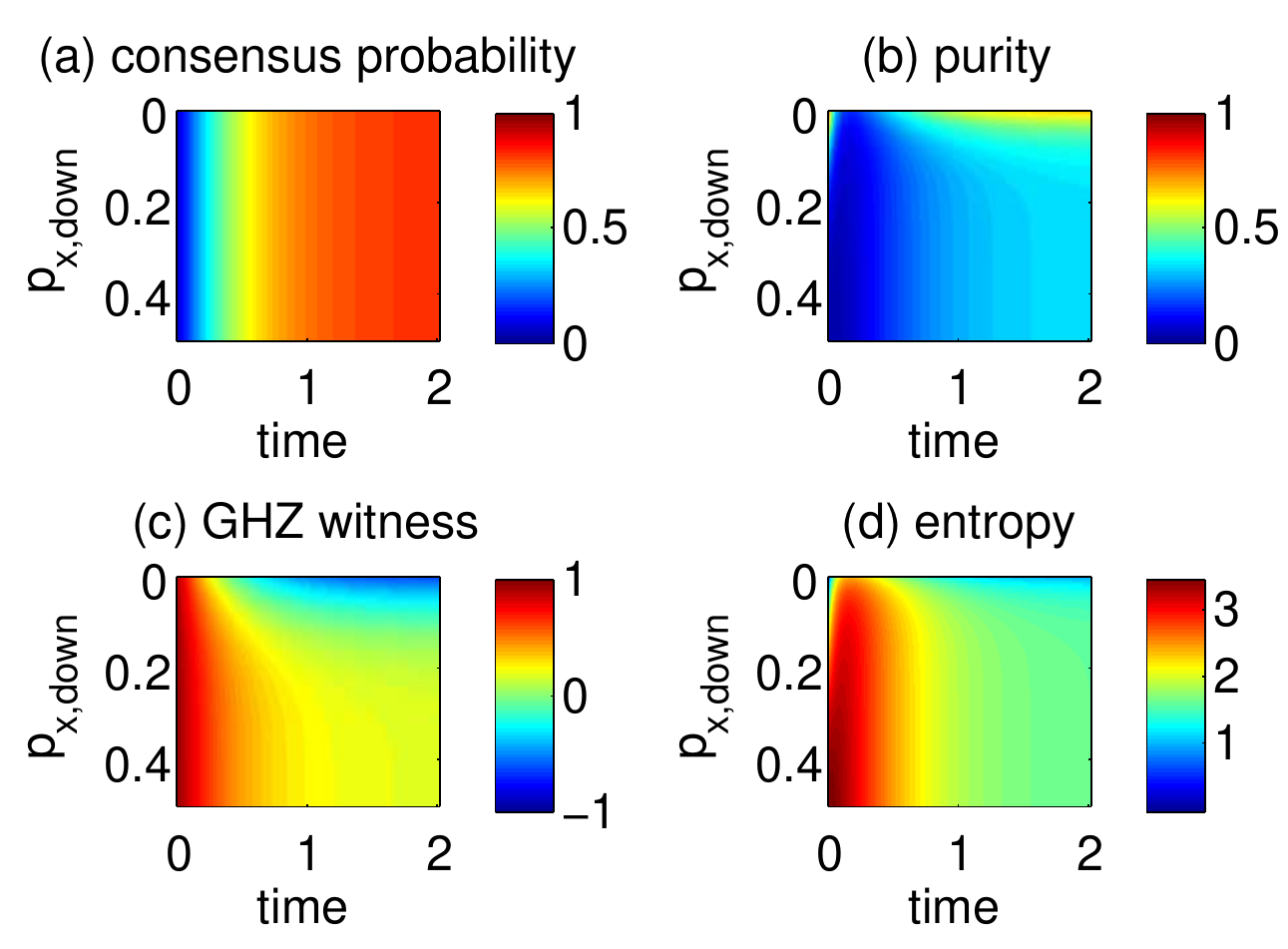}
 \caption{(Color online) Quantum consensus dynamics for the initial state
$[(1-p_{x,\text{down}})\ket{+}\bra{+} +p_{x,\text{down}}\ket{-}\bra{-}]^{\otimes N}$.
Time is measured in $\Gamma_{\text{copy}}^{-1}$. Here, $\Gamma_{\text{flip}}=0.1\Gamma_{\text{copy}}$.}
 \label{fig:init-pd}
\end{figure}

Figs.~\ref{fig:init-theta}-~\ref{fig:init-pd} show that GHZ-entanglement is also generated even for these two types of initial states, if the deviation is sufficiently small.
Fig.~\ref{fig:init-theta} shows that $W_{\text{GHZ}}<-0.5$ is achieved at the stationary value when $\theta<0.1 \pi $.
Fig.~\ref{fig:init-pd} shows that $W_{\text{GHZ}}<-0.5$ is achieved at the stationary value when $p_{x,\text{down}}<0.1 $.

\section{Copy processes by classical feedback controls}
\label{apx:clsFB}

Here we show another type of feedback control (which we called classical feedback control in the main text), which also assimilates the copy processes.

For the copy process that the qubit $i$ copies $j$, the feedback controller first measures all four possible states $(s_i,s_j) = (0,0), (0,1), (1, 0), (1,1)$ of the qubits.
Let the measurement outcomes be $k=0,\cdots 3$, respectively.
The measurements are described by the projection operators
$\tilde{M}_0^{(ij)} = \ket{00}\bra{00}$, $\tilde{M}_1^{(ij)} = \ket{01}\bra{01}$, $\tilde{M}_2^{(ij)} = \ket{10}\bra{10}$, and $\tilde{M}_3^{(ij)} = \ket{11}\bra{11}$.
Then, according to the measurement outcome, the feedback operations
$\tilde{U}_0^{(i)} = \mathbb{1} $, $\tilde{U}_1^{(i)} = X_i $, $\tilde{U}_2^{(i)} = X_i$, and $\tilde{U}_3^{(i)} = \mathbb{1} $ occurs.
The copy process maps the qubits from $\rho$ to $\tilde{C}_{i \leftarrow j}(\rho)$
\begin{equation}
 \label{eq:tildeC}
 \tilde{C}_{i \leftarrow j}(\rho)= \sum_{k=0,\cdots, 3} \tilde{U}_k^{(i)} \tilde{M}_k^{(ij)} \rho \tilde{M}_k^{(ij)} \big(\tilde{U}_k^{(i)}\big)^\dagger.
\end{equation}

\begin{figure}[h]
 \centering \includegraphics[width=0.85\columnwidth]{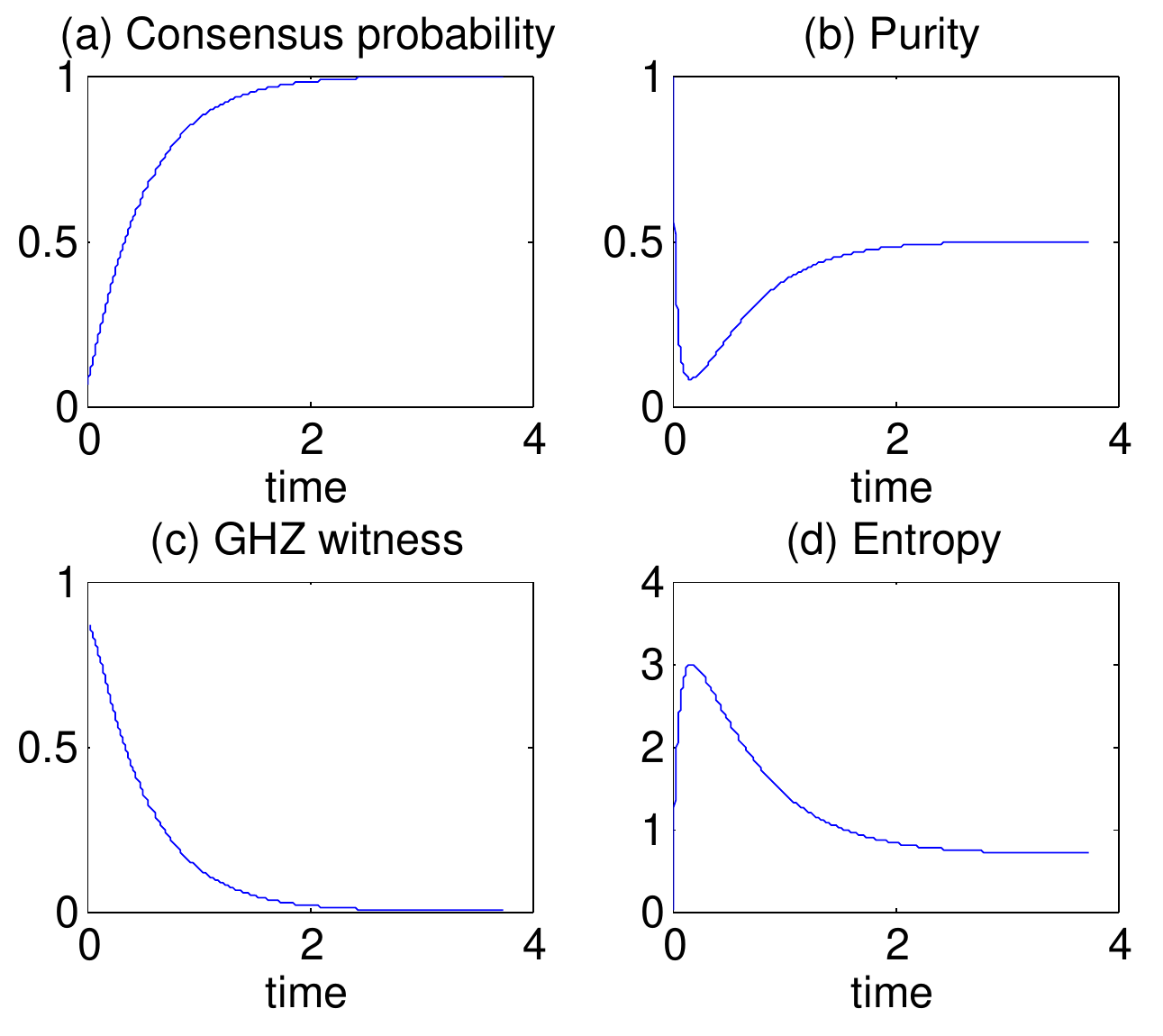}
 \caption{Consensus dynamics by classical feedback controls in comparison with Fig.~\ref{fig:ghzN} of the main text. The time is measured in $\Gamma_{\text{copy}}^{-1}$.}
 \label{fig:clsFB}
\end{figure}

Fig.~\ref{fig:clsFB} shows the consensus dynamics by $\tilde{C}_{i\leftarrow j}$ in the absence of noise and the initial state of the symmetric superposition, situation of Fig.~\ref{fig:ghzN}.
Although the consensus probability reaches the value $1$, the GHZ entanglement is not generated as $W_{\text{GHZ}} \rightarrow 0$.
This is because the state approaches the classical ensemble $\ket{00}\bra{00} + \ket{11}\bra{11})$, as evidentiated by the consensus probability of $1$ and purity of $0.5$. The decoherence appears as the measurements of the feedback control $\tilde{C}_{i \leftarrow j}$ destroy the coherence of the qubits $i$ and $j$ in the opinion basis $\{ \ket{0_i0_j}, \ket{0_i1_j}, \ket{1_i0_j}, \ket{1_i1_j}\}$.

\section{Upper bound of the entropy reduction rate}
\label{sec:ubEnt}
Here we derive the upper bound of the entropy reduction rate by the entangling Maxwell demon, $-\dot{S}_{\text{copy}}$. To this purpose, we first formulate the integral quantum fluctuation theorem (IQFT) for the entropy change, see Eq.~(\ref{eq:iQFT}). The formulation is based on the approach of Ref.~\cite{funo2015quantum} on the IQFT for a single quantum feedback control. Then, by applying Jensen's inequality to the obtained IQFT, we derive the upper bound for the entropy reduction rate, Eq.~(\ref{eq:dSdt-lambda}).
The notations in the derivations are presented in the case of the quantum voter model, but they are straightfowardly generalized for a generic entangling Maxwell demon.

To obtain the IQFT, let us consider a thought experiment monitoring the time evolution during a small time $\Delta t$ which is much smaller than the repetition period $\Gamma_{\text{copy}}^{-1}$ but larger than the operation time of a feedback control (i.e. the time for completing three C-NOT gates in the case of the quantum voter model).\\
i) The initial state $\rho$ is measured in the basis $\{\ket{x}\}$ which diagonalizes $\rho$.\\
ii) We monitor, if any, which particles were selected for a quantum feedback control.\\
If a pair of qubits $i$ and $j$ are selected,\\
iii) we monitor the measurement outcome $k$ (which is 0 when $s_i = s_j$ and 1 otherwise) of the feedback control applied to the selected qubits. \\
iv) We measure the post-measurement state in the basis $\{\ket{y}\}$ which diagonalizes $\rho_k^{(ij)}$.\\
v) We measure the feedback-operated state in the basis $\{\ket{z}\}$ that diagonalizes a reference state~\cite{funo2015quantum} $\rho_r$.\\
A quantum transition is described by the record of all the monitoring $(x,ij, k, y,z)$.
The corresponding transition probability $P(x, ij, k, y,z)$ is 
product of the probabilities of each monitoring i) -- v),
\begin{align}
 P(x,ij,k,y,z)
 &= p_{\text{i}}(x) p_{\text{copy}} p_{\text{iii}}(k|x,ij) p_{\text{iv}}(y|x,ij,k) \nonumber \\
 &\quad \times p_{\text{v}}(z|ij,k,y) + \mathcal{O}(\Gamma_{\text{copy}} \Delta t)^2, \label{eq:P} \\
 p_{\text{i}}(x) &= \braket{x|\rho|x} , \\
 p_{\text{copy}} &= \Gamma_{\text{copy}}\Delta t ,\\
 p_{\text{iii}}(k|x,ij) &= |\braket{x|M^{(ij)}_k |x}|^2 , \\
 p_{\text{iv}}(y|x,ij,k) &= |\braket{y| M^{(ij)}_k |x}|^2/p_{\text{iii}}(k|x,ij) , \\
 p_{\text{v}}(z|ij,k,y) &= |\braket{z|U_k^{(i)}|y}|^2 .
\end{align}
The transition probability $P$ is related to the probability $P^{(ij)}$ given that the particles $i$ and $j$ are selected,
\begin{align}
 P(x, ij, k, y, z) &= p_{\text{copy}} P^{(ij)}(x,k,y,z)
 + \mathcal{O}(\Gamma_{\text{copy}}\Delta t)^2, \label{eq:P-Pij} \\
 P^{(ij)}(x,k,y,z)&= p_{\text{i}}(x) p_{\text{iii}}(k|x,ij) p_{\text{iv}}(y|x,ij,k) p_{\text{v}}(z|ij,k,y), 
\end{align}

Following the approach of Ref.~\cite{funo2015quantum}, we define the unaveraged entropy change $\sigma$ by the feedback controls and unaveraged
quantum-classical mutual information $i_{\text{QC}}$~\cite{funo2015quantum} for a transition $(x,ij, k, y,z)$,
\begin{align}
 \sigma(x,z) &= \ln p_{\text{i}}(x) -\ln p_r(z), \label{eq:sig}\\
 i_{\text{QC}}(x,ij,k,y) &= -\ln p_{\text{i}}(x) +\ln p(y|ij,k). \label{eq:iQC}
\end{align}
Here $p_r(z) \equiv \braket{z|\rho_r|z}$ is the probability distribution of the reference state, and $p(y|ij,k)=\braket{y|\rho_k^{(ij)}|y}$ is the probability distribution of the post-measurement state $\rho_k^{(ij)}$.
If no feedback control has occurred, $\sigma= 0$ and $i_{\text{QC}}=0$ because we want to describe entropy reduction by the feedback controls.

The expectation value of $\sigma$ provides an lower-bound for the entropy change by the feedback controls, $\dot{S}_{\text{copy}}$ [see Eq.~(\ref{eq:dS-copy})], during the time $\Delta t$,
\begin{align}
 \braket{\sigma}
 &= N_{\text{copy}} \Gamma_{\text{copy}}
 \big\{S(C(\rho)) - S(\rho) \big\} \Delta t \label{eq:exp-sig} \\
 &\le \dot{S}_{\text{copy}} \Delta t , \label{eq:sig-Scopy}
\end{align}
when choosing the reference state as the result of the stochastic feedback control without knowing which particles were selected,
\begin{equation}
 \label{eq:rho-r-apx}
 \rho_r = C(\rho).
\end{equation}
Eq.~(\ref{eq:exp-sig}) is derived by relating $\braket{\sigma}$ to a expectation value $\braket{\sigma}^{(ij)}$ given that the particles $i$ and $j$ were selected, and using that $\braket{\sigma}^{(ij)}= -S(\rho) -\text{Tr}[ C_{i \leftarrow j}(\rho) \ln C(\rho)]$ (see Ref.~\cite{funo2015quantum}).
\begin{align}
 \braket{\sigma}
 &= \sum_{i\neq j} p_{\text{copy}} \braket{\sigma}^{(ij)} 
 \nonumber \\
 &=\sum_{i \neq j} p_{\text{copy}} \Big[- S(\rho)
 -\text{Tr} \big[ C_{i \leftarrow j}(\rho) \ln C(\rho) \big] \Big] \nonumber \\
 &= - N_{\text{copy}} p_{\text{copy}} S(\rho)
 -N_{\text{copy}} p_{\text{copy}} \text{Tr} \big[ C(\rho) \ln C(\rho) \big] \nonumber \\
 &= N_{\text{copy}} p_{\text{copy}} [-S(\rho)+S(C(\rho))] \nonumber\\
 &= N_{\text{copy}} \Gamma_{\text{copy}} [-S(\rho)+S(C(\rho))]\Delta t.
\end{align}
Eq.~(\ref{eq:sig-Scopy}) is derived when applying the concavity of the von Neumann entropy to the definition of $\dot{S}_{\text{copy}}$,
\begin{equation}
 \dot{S}_{\text{copy}}
 = \frac{S\Big( N_{\text{copy}} p_{\text{copy}} C(\rho) + (1-N_{\text{copy}}p_{\text{copy}}) \rho \Big) -S(\rho) }{\Delta t} .
\end{equation}

The expectation value of $i_{\text{QC}}$ is the average quantum-classical mutual information obtained in the feedback controls during the time $\Delta t$, 
\begin{align}
 \braket{i_{\text{QC}}}
 &=\sum_{i \neq j} p_{\text{copy}} I_{\text{QC}}^{(ij)}, \\
 &= N_{\text{copy}}\Gamma_{\text{copy}} \overline{I_{\text{QC}}} \Delta t. \label{eq:exp-iQC}
\end{align}

\begin{figure}[h]
 \centering
 \includegraphics[width=\columnwidth]{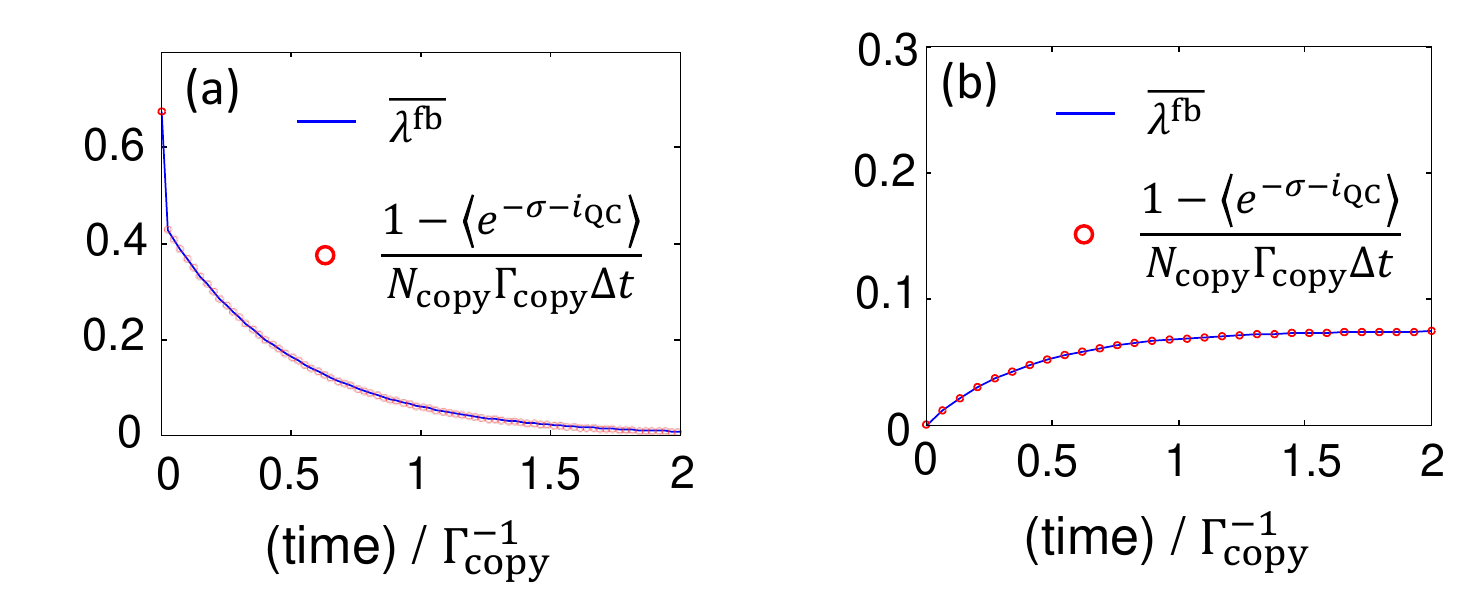}
 \caption{Numerical verification of Eq.~(\ref{eq:iQFT}). The case (a) corresponds to the case of Fig.~\ref{fig:S-I}(a)--(b), and (b) corresponds to Fig.~\ref{fig:S-I} (c)--(d). Here, $\Delta t = 0.0025 \Gamma_{\text{copy}}^{-1} $ for (a) and $\Delta t = 0.01 \Gamma_{\text{copy}}^{-1} $ for (b).}
 \label{fig:FT}
\end{figure}

To obtain the integral quantum fluctuation theorem, we express $\braket{e^{-\sigma-i_{\text{QC}}}}$ by an expectation value $\braket{e^{-\sigma-i_{\text{QC}}}}^{(ij)}$ given that the particles $i$ and $j$ were selected.
We use the integral fluctuation theorem for a single feedback control~\cite{funo2015quantum},
\begin{equation}
\braket{e^{-\sigma-i_{\text{QC}}}}^{(ij)}= 1-\lambda_{\text{fb}}^{(ij)},
\label{eq:IQFT-1fb}
\end{equation}
and use that if no feedback control has occurred, $\sigma=0$ and $i_{\text{QC}}=0$. Then, we obtain 
\begin{align}
 \braket{e^{- \sigma -i_{\text{QC}}}}
 &= \sum_{i \neq j} p_{\text{copy}}
 \braket{e^{ -\sigma-i_{\text{QC}}}}^{(ij)}
 + (1-\sum_{i \neq j}p_{\text{copy}}) \nonumber \\
 &= \sum_{i \neq j} p_{\text{copy}} (1-\lambda^{(ij)}_{\text{fb}})
 + (1-N_{\text{copy}} p_{\text{copy}}), \nonumber \\
 &= 1 -N_{\text{copy}} p_{\text{copy}} \overline{\lambda_{\text{fb}}}.
 \label{eq:IQFT-sup}
\end{align}
This leads to the integral quantum fluctuation theorem during the small time $\Delta t \ll \Gamma_{\text{copy}}^{-1}$ (note that Eqs.~(\ref{eq:P}) and (\ref{eq:P-Pij}) are only valid for such small time)
 \begin{equation}
 \label{eq:iQFT}
 \braket{e^{-\sigma -i_{\text{QC}}}}
 = 1 -N_{\text{copy}}\Gamma_{\text{copy}}\Delta t \overline{\lambda_{\text{fb}}} + \mathcal{O}(\Gamma_{\text{copy}}\Delta t)^2.
\end{equation}
Fig.~\ref{fig:FT} shows that Eq.~(\ref{eq:iQFT}) is indeed satisfied for the situations of Fig.~\ref{fig:S-I}.

Finally, we obtain the upper bound of the entropy reduction rate $-\dot{S}_{\text{copy}}$ by applying Jensen's inequality, $e^{\braket{X}}\le\braket{e^X}$, to Eq.~(\ref{eq:iQFT}) and taking the limit of $\Gamma_{\text{copy}} \Delta t \rightarrow 0$,
\begin{align}
 \dot{S}_{\text{copy}} &\ge 
 \frac{\braket{\sigma}}{\Delta t} \nonumber \\
 &\ge \frac{1}{\Delta t} \Big[ -\braket{i_{\text{QC}}} 
 -\ln \big[ 1- N_{\text{copy}} \Gamma_{\text{copy}} \Delta t \big] \overline{\lambda_{\text{fb}}} \Big] \nonumber \\
 &= N_{\text{copy}} \Gamma_{\text{copy}}
 \big[- \overline{I_{\text{QC}}} + \overline{\lambda_{\text{fb}}}\big] \label{eq:sig-LB} 
\end{align}
In the last equality, we used Eq.~(\ref{eq:exp-iQC}) and that $\Gamma_{\text{copy}} \Delta t$ is small.

\section{Absolute irreversibility in the quantum voter model for the symmetric initial state}
\label{apx:lambda-symm}

Here we derive that the absolute irreversibility $\overline{\lambda_{\text{fb}}}$ in the quantum voter model for
the symmetric initial state $\rho = (\sum_{s_1,\cdots, s_N =0,1} \ket{s_1 \cdots s_N}/\sqrt{2^N})(\text{h.c.})$ 
\begin{equation}
 \label{eq:lambda-symm}
\overline{\lambda_{\text{fb}}}= \frac{3}{4} \left[1-\frac{2}{N(N-1)} \right].
\end{equation}

 We use Eq.~(\ref{eq:lambda-pure-S2}). We evaluate the overlap $ \text{Tr} [ C_{i \leftarrow j}(\rho) C_{l \leftarrow m}(\rho)]$ by dividing the choices of $i,j,l$ and $m$ in three cases.

i) When the qubit pair $(i,j)$ does not share any common qubit with the pair $(l,m)$, the overlap $ \text{Tr} [ C_{i \leftarrow j}(\rho) C_{l \leftarrow m}(\rho)]$ is 1/4. This is because the overlap is the same as $ \text{Tr} [ C_{1 \leftarrow 2}(\rho) C_{N-1 \leftarrow N}(\rho)]$ due to the permutation symmetry of qubits, and it equals $|\braket{\Phi_+ +\cdots + | + \cdots + \Phi_+}|^2= |\braket{\Phi_+++| ++\Phi_+}|^2 = 1/4 $.

ii) When the qubit pair $(i,j)$ share one common qubit with the pair $(l,m)$, the overlap $ \text{Tr} [ C_{i \leftarrow j}(\rho) C_{l \leftarrow m}(\rho)]$ is also 1/4.
This is because the overlap equals $ \text{Tr} [ C_{1 \leftarrow 2}(\rho) C_{2 \leftarrow 3}(\rho)]$ due to the permutation symmetry, and it equals $|\braket{\Phi_+ +\cdots + |+\cdots + \Phi_+}|^2= |\braket{\Phi_++| +\Phi_+}|^2 = 1/4 $.

iii) When the qubit pair $(i,j)$ share two common qubits with the pair $(l,m)$, the overlap $ \text{Tr} [ C_{i \leftarrow j}(\rho) C_{l \leftarrow m}(\rho)]$ is 1, due to the permutation symmetry.

We count the number of the choices of $i,j,l$, and $m$ in the cases i) -- iii).
The total number of choices in all the cases is $[N(N-1)]^2$, due to the all-to-all network connectivity, $i\neq j$ and $l \neq m$.
The number of choices in the case iii) is $2N(N-1)$. 
The number of choices in the cases i) and ii) is $[N(N-1)]^2 -2N(N-1)$.
Using these in Eq.~(\ref{eq:lambda-pure-S2})
\begin{align}
 \overline{\lambda_{\text{fb}}}
 &= \frac{[N(N-1)]^2 -2N(N-1) }{N_{\text{copy}}^2}(1-\frac{1}{4}) \nonumber \\
 & \qquad
 + \frac{2N(N-1) }{N_{\text{copy}^2}}(1-1).
\end{align}
Replacing $N_{\text{copy}}= N(N-1)$, this is equivalent to Eq.~(\ref{eq:lambda-symm}).

\section{Steady-state solution for the two-qubit information engine}
\label{apx:SSheatEngine}

Here we derive the steady-state solution near the maximally entangled state in situation of the two-qubit information engine discussed in Sec.~\ref{sec:egDemon}.
We try the following ansatz
\begin{equation}
 \label{eq:2qb-SS}
 \rho_{\text{st}} = (1-\epsilon) \rho_\text{C} + \epsilon \rho_{\text{FC}},
\end{equation}
to which we apply the steady-state condition $\dot{\rho}_\text{st}=0$ with the master equation Eq.~(\ref{eq:L-2qbs}).
The unitary part of the Liouvillian vanishes,
\begin{equation}
 [H, \rho_{\text{st}}]=0 ,
\end{equation}
because
\begin{align}
 Z_1\otimes Z_2 \, \rho_\text{C}\, Z_1\otimes Z_2 &= \rho_\text{C}, \\
 Z_1\otimes Z_2 \, \rho_\text{FC}\, Z_1\otimes Z_2 &= \rho_\text{FC} .
\end{align}

To calculate the other parts of the Liouvillian, we use
\begin{align}
 C_{i\leftarrow j}(\rho_\text{C}) &= \rho_\text{C} , \\
 C_{i\leftarrow j}(\rho_\text{FC}) &= \rho_\text{C} , \\
 X_i \rho_\text{C} X_i &= \rho_\text{FC} , \\
 X_i \rho_\text{FC}X_i &= \rho_\text{C} ,
\end{align}
 for any $i$ and $j\neq i$.
These lead to 
\begin{align}
 &C_{i \leftarrow j}(\rho_\text{st}) -\rho_\text{st}
 = \epsilon (\rho_\text{C} -\rho_\text{FC}), \\
 &X_i \rho_\text{st}X_i -\rho_\text{st}
 = (2\epsilon-1) (\rho_\text{C} -\rho_\text{FC}), \label{eq:XrX}\\
 &\dot{\rho}_\text{st} = (\rho_\text{C} -\rho_\text{FC} )
 [ 2 \epsilon \Gamma_{\text{copy}} + 2 ( 2 \epsilon -1) \Gamma_{\text{flip}} ].
\end{align}
Therefore, the steady-state condition is satisfied when
\begin{equation}
 \label{eq:epsilon}
 \epsilon = \frac{\Gamma_{\text{flip}}}{\Gamma_{\text{copy}} + 2 \Gamma_{\text{flip}}}.
\end{equation}

\bibliographystyle{plainnat}

\bibliography{QV}

\end{document}